\documentstyle[12pt]{article}
\def\fcskp{\baselineskip 12pt}              
\newcommand{\fc}{\footnotesize \fcskp \advance\itemsep by -6pt}
\def\DREZ{\hbox{${\rm DR\overline{EZ}}$}}
\def\DRED{{\rm DRED}}
\def\NDR{{\rm NDR}}
\def\HV{{\rm HV}}
\def\CONT{{\em cont}}
\def\LATT{{\em lat}}

\def\SM{{\rm SM}}
\def\FSM{{\rm F,SM}}
\def\FPR{{\rm F,PR}}
\def\tM{{\widetilde M}}
\def\half{{\textstyle{1\over2}}}

\def\fourth{{\textstyle{1\over4}}}
\def\frac#1#2{{\textstyle{#1\over#2}}}

\def\Tr{\mathop{\rm Tr}\nolimits}
\def\bar#1{\overline{#1}}
\def\chibar{\overline\chi}

\def\vev#1{\langle #1 \rangle}
\def\cm{{\cal M}}
\def\cmb{\bar{{\cal M}}}
\def\co{{\cal O}}
\def\cp{{\cal P}}

\def\ca{{\cal A}}

\def\vdag{\vphantom{\dag}}
\def\g#1{\gamma_{#1}}
\def\gd#1{\gamma_{#1}^{\dag}}

\def\gam#1{\overline{(\gamma_{#1}\otimes 1)} }
\def\ggam#1{\overline{\overline{(\gamma_{#1}\otimes 1)} } }

\def\iiden{\overline{\overline{(1\otimes 1) } } }
\def\sfno#1#2{\overline{(\gamma_{#1}\otimes\xi_{#2})}}

\def\sf#1#2#3#4{\overline{(\gamma_{#1}\otimes\xi_{#2})}_{#3#4}}
\def\ssf#1#2#3#4{\overline{\overline{(\gamma_{#1}\otimes\xi_{#2})}}_{#3#4}}
\def\operii{(1\otimes1)}
\def\operix#1{(1\otimes\xi_{#1})}
\def\opergi#1{(\g{#1}\otimes1)}
\def\opergx#1#2{(\g{#1}\otimes\xi_{#2})}
\def\sbar{\bar{s}}
\def\cbar{\bar{c}}
\begin{document}
\begin{titlepage}
 \null
 \begin{center}
 \makebox[\textwidth][r]{UW/PT-92-13}
 \makebox[\textwidth][r]{CEBAF-TH-92-20}
 \par\vspace{0.25in} 
  {\Large
	PERTURBATIVE CORRECTIONS FOR STAGGERED \\
        FERMION BILINEARS}
  \par
 \vskip 2.0em
 {\large
  \begin{tabular}[t]{c}
	Apoorva Patel \footnotemark\\[1.em]
	\em CTS and SERC, Indian Institute of Science\\
	\em Bangalore, 560012, India\\[1.5em]
	Stephen R. Sharpe \footnotemark\\[1.em]
	\em Continuous Electron Beam Accelerator Facility \\
	\em Newport News, VA 23606 \\
	{and}\\
	\em Physics Department, FM-15, University of Washington \\
	\em Seattle, WA 98195 \\
  \end{tabular}}
 \par \vskip 3.0em
 {\large\bf Abstract}
\end{center}
\quotation
We calculate the perturbative corrections to
fermion bilinears that are used in numerical simulations
when extracting weak matrix elements using staggered fermions.
This extends previous calculations of Golterman and Smit,
and Daniel and Sheard.
In particular, we calculate the corrections for non-local bilinears
defined in Landau gauge with gauge links excluded.
We do this for the simplest operators,
i.e. those defined on a $2^4$ hypercube,
and for tree level improved operators which live on $4^4$ hypercubes.
We also consider gauge invariant operators in which the ``tadpole''
contributions are suppressed by projecting the sums
of products of gauge links back in to the gauge group.
In all cases, we find that the variation in the size of the perturbative
corrections is smaller than those with the gauge invariant unimproved
operators. This is most strikingly true for the smeared operators.
We investigate the efficacy of the mean-field method of Lepage
and Mackenzie at summing up tadpole contributions.
In a companion paper we apply these results to four-fermion operators.
\endquotation


\footnotetext[1]{Email: adpatel@cts.iisc.ernet.in}
\footnotetext{Email: sharpe@galileo.phys.washington.edu}
\vfill
\mbox{October 1992}
\end{titlepage}

\section{INTRODUCTION}
\label{sintro}
Lattice calculations of hadronic matrix elements are beginning
to have a significant impact on phenomenology \cite{lusignoli}.
A necessary step in such calculations is the determination of
the relationship between the continuum operators whose matrix
elements we are interested in, and the lattice operators whose
matrix elements we can compute.
The matrix elements of lattice and continuum operators
differ due to contributions involving loop momenta close to the cut-off,
i.e. $p \sim \pi/a$,
where $a$ is the lattice spacing.
As long as $a$ is small enough ($1/a$ is $2-4$ GeV in present calculations),
the difference can be calculated using perturbation theory.
In the present work we extend such calculations to the staggered fermion
operators which are used
in the most accurate numerical simulations \cite{bkprl,sharpelat91}.

Perturbative calculations with staggered fermions were pioneered by
Sharatchandra, Thun and Weisz \cite{sharatchandra},
extended to a variety of fermion bilinears by Golterman and Smit
\cite{goltermansmit},
and extended again to all bilinears by Daniel and Sheard \cite{daniel}.
We refer to the latter two papers as GS and DS respectively.
The present calculation is an extension of the work of DS;
we follow their notation, and use and extend the technology they developed.
We recommend that readers unfamiliar with the subject consult the
above mentioned references for the technical background.

A single staggered lattice fermion corresponds to $N_f=4$ Dirac fermions
in the continuum limit, due to the existence of sixteen ``doublers''.
This has two important consequences.
First, the continuum theory to which one must match the operators
has $N_f$ degenerate flavors for each type of staggered fermion.
Second, to construct local bilinears having all possible spins and
flavors one must use lattice operators in which the quark and antiquark
fields are spread out over a $2^4$ hypercube \cite{kluberg}.
Since quark and antiquark fields are, in general, on different sites,
one must make the operators gauge invariant.
The most straightforward method is to insert the average of
the products of gauge links along the paths of minimal length.
These were the operators considered in DS, but they have not,
in general, been used in numerical simulations.\footnote{%
Two recent calculations have, however, used these operators
\cite{horsley,tsukuba}.}
This is in part because they have large
perturbative corrections due to the fluctuating gauge links.
The corrections can, however, be reduced substantially by summing up
the tadpole diagrams \cite{sheardtad,lepagemackenzie}.
We discuss this issue in detail in the last section.

The first simulations attempting to extract matrix elements of such
operators used operators in which the average of the product of gauge links
is projected back in to the gauge group \cite{wius}.
The projection reduces the fluctuations to the
level of that of a single link.
We have calculated the corrections for these operators.
This turns out to require only a slight modification of the work of DS.

Most recent calculations (e.g. Ref. \cite{bkprl}) use operators in which
the lattices are fixed to Landau gauge and the gauge links are dropped.
This removes the problem of fluctuating gauge links,
and also makes the operators much more simple to use in practice.
The only difficulties are the existence of Gribov copies in Landau gauge,
and the fact that, formally,
one cannot define a transfer matrix when using such operators.
One can argue, however, that these problems do not have a significant
effect on the numerical results, at least in the continuum limit \cite{hmks}.
In any case, these are not problems in perturbation theory,
in which Landau gauge is uniquely defined.
The extension of the calculation of DS to these operators is
presented below.
A side benefit of the calculation is that we can check that,
for gauge invariant operators, the results in Landau gauge
agree with those in Feynman gauge.

A major problem with the matrix elements calculated using
Landau gauge operators is that there are large $O(a)$ corrections
\cite{sharpelat90,sharpelat91}.\footnote{%
Matrix elements of gauge invariant operators also have $O(a)$ corrections.
These appear to be of similar size to those of the Landau gauge operators
\cite{tsukuba}.
We focus on Landau gauge operators since they are simpler to improve.}
With staggered fermions the action is accurate to $O(a^2)$,
so to remove such corrections one need only improve the operators.
It is straightforward to improve the Landau gauge operators at tree level,
and a simple method for doing this has been proposed in Ref. \cite{book}.
It requires smearing the quark and antiquark fields in such a way
that the entire operator gets spread out over a $4^4$ hypercube.
We have calculated the perturbative corrections for these improved operators,
and we are also using them in numerical simulations \cite{sharpelat91}.

Many of the phenomenologically interesting matrix elements
involve four-fermion operators rather than bilinears.
Again, numerical simulations have used Landau gauge four-fermion operators.
For these, one can, as for Wilson fermions \cite{mart4fermion},
obtain the perturbative corrections from those for bilinears.
The main difficulty is that Fierz transformations must account
for the extra flavors associated with staggered fermions.
We present the results in a separate publication \cite{usinprep}.
The calculation for gauge invariant four-fermion operators,
which requires additional work, was done some time ago for the
left-left chiral operators by Sheard \cite{sheard}.
It has recently been repeated by Ishizuka and Shizawa \cite{tsukubapert}.
Results for the left-right penguin operators are presented
in our companion paper \cite{usinprep}.

The outline of this paper is as follows. In Section \ref{snotation}
we introduce the notation and give the Feynman rules.
Section \ref{sdetails} gives the details of the computations.
We collect the results in Section \ref{sresults},
and close in Section \ref{sconcl} with a discussion of the results.

The results of this work have been presented in preliminary form
in Refs. \cite{ssringberg} and \cite{book}. Various minor errors
in these earlier presentations are corrected herein.
Ishizuka and Shizawa have independently calculated the corrections
for the gauge invariant and unsmeared Landau gauge operators
\cite{tsukubapert},
with results that are in complete agreement with ours.

\section{NOTATION AND DEFINITIONS}
\label{snotation}
In this section we present the definitions and
the Feynman rules necessary for the subsequent calculation.
A number of the results can be read directly from DS,
and we do not repeat those here.
Other results we have taken from DS but re-expressed so as to
highlight the similarity of the lattice Feynman rules
to those of a continuum theory with four flavors.
Finally, the rules relating to the use of Landau gauge
and the smeared operators are new.

\subsection{Dirac matrices}

Following DS, we use a gamma matrix basis for both spin and flavor.
To enumerate this basis we use ``hypercube vectors'',
four-vectors whose components are 0 or 1.
These vectors are combined using modulo-2 operations,
which, if confusion is likely to arise,
we denote with a subscript 2 (e.g. $=_2$, $+_2$).
A general gamma matrix is labeled by a hypercube vector
\begin{equation}
  \g{S} = \g{1}^{S_1} \g{2}^{S_2} \g{3}^{S_3} \g{4}^{S_4} \ .
\end{equation}
We use Euclidean space gamma matrices, which are hermitian, and
satisfy $\{\g\mu,\g\nu\}=2\delta_{\mu\nu}$.
An alternative basis is built out of the complex conjugate matrices
$\xi_\mu=\g\mu^*$, and is labeled similarly
\begin{equation}
  \xi_F = \xi_1^{F_1} \xi_2^{F_2} \xi_3^{F_3} \xi_4^{F_4} \ .
\end{equation}
In the products of such matrices we use the notation
\begin{equation}
  \g{S}  \g{S'}   = \g{SS'}\ , \quad
  \xi_{S}\xi_{S'} = \xi_{SS'} \ .
\end{equation}
Useful notation for hypercube vectors is
\begin{equation}
  S\cdot F = \sum_\mu S_\mu F_\mu \ ,\quad
  \widetilde{S}_\mu =_2 \sum_{\nu\ne\mu} S_\mu \ .
\end{equation}
In manipulating products of gamma matrices we make frequent use of
the following relations
\begin{eqnarray}
\label{gettildeeqn}
  \g{S}\g{F}\gd{S} &=&(-)^{S\cdot \widetilde{F}} \g{F}
                  \ =\ (-)^{\widetilde{S}\cdot F} \g{F}\ ,\\
  \g{\mu}\g{S} &=&\g{[\mu+S]}\ \eta_\mu(S) \ ,\quad
  \eta_\mu(S) \ =\ (-)^{\sum_{\nu<\mu} S_\nu} \ , \\
  \g{S}\g{\mu} &=&\g{[\mu+S]}\ \zeta_\mu(S)\ ,\quad
  \zeta_\mu(S)\ =\ (-)^{\sum_{\nu>\mu} S_\nu} \ .
\end{eqnarray}

\subsection{Continuum bilinears}

We label quark fields in a continuum theory with four degenerate
fermions using upper case letters, e.g. $Q_{\alpha,a}$.
These have a spinor index (here $\alpha$) and a flavor index
(here $a$), both running from 1 to 4.
Color indices play no role in the discussion of this section, and we
do not show them explicitly.
A general bilinear is specified by a spin and a flavor matrix
\begin{equation}
  \co_{SF}^\CONT =
  \bar Q_{\alpha,a} \g{S}^{\alpha\beta} \xi_F^{ab} Q_{\beta,b} \ .
\end{equation}
To keep the notation as clear as possible,
we always use $\gamma$ matrices for spin and $\xi$ matrices for flavor.
It is convenient to combine spin and flavor matrices in to a single
$16\times16$ matrix $\opergx SF$
\begin{equation}
  \co_{SF}^\CONT
  \ = \     \bar Q_{\alpha,a} \opergx{S}{F}^{\alpha a,\beta b} Q_{\beta,b}
  \ \equiv\ \bar Q \opergx{S}{F} Q \ .
\end{equation}
The second form is a useful abbreviation in which
we treat $Q$ as a 16-component column vector.
For example, the scalar-isoscalar bilinear $\bar Q_{\alpha,a} Q_{\alpha,a}$
is written as $\opergx SF$ with $S=F=(0000)$, or equivalently as $\operii$.

We also need two other sets of matrices
which are unitarily equivalent to $\opergx SF$.
Both sets trade the indices $\{ \alpha,a \}$ for a hypercube vector
\cite{kieu,daniel} :
\begin{eqnarray}
  \sf S F A B &\equiv&
  \frac14 \Tr[ \g{A}^{\dag} \g{S}^{\vdag} \g{B}^{\vdag} \g{F}^{\dag} ] \\
  \label{hypercuberep}
  &=&\sum_{\alpha,a,\beta,b}      (\frac12\g{A})^{\alpha a}
  \opergx{S}{F}^{\alpha a,\beta b} (\frac12\xi_{B})^{\beta b} \ ; \\
  \label{polerep}
  \ssf S F A B &\equiv&
  \sum_{CD} \frac14 (-)^{A.C}\ \sf S F C D \ \frac14 (-)^{D.B} \ .
\end{eqnarray}
We refer to the former set as being in the ``hypercube'' basis,
while the latter we say is in the ``momentum'' basis.
The ``hypercube'' basis is convenient for numerical simulations,
while the ``momentum'' basis is convenient for perturbative calculations.
The multiplication rule for each of the representations is
as for a direct product, for example
\begin{equation}
  \sfno{S}{F}\ \sfno{S'}{F'} =
  \overline{(\g{S}\g{S'} \otimes \xi_{F}\xi_{F'})}
  \equiv \sfno{SS'}{FF'} \ .
\end{equation}

\subsection{Feynman rules}

\begin{figure}
\vspace{1.truein}
\includegraphics{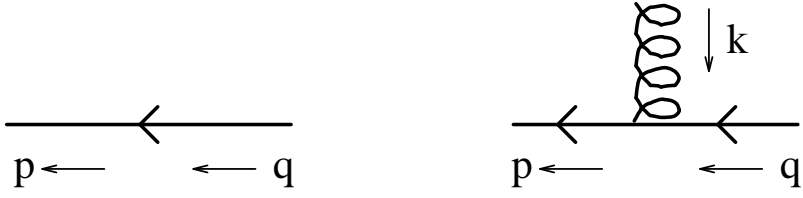}
\bigskip
\caption[feynmanfig]{\fc
Notation for Feynman diagrams.}
\label{feynmanfig}
\end{figure}

Since continuum fermions are constructed from the neighborhood of the
16 lattice poles at $p_\mu=\pi A_\mu$,
it is convenient to use the decomposition
\begin{equation}
  p_\mu = p'_\mu + \pi A_\mu \ ;\qquad -\pi/2 < p'_\mu \le \pi/2 \ .
\end{equation}
For physical quarks the ``small'' component $p'$ is close to zero:
if $p_{\rm phys}$ is the physical momentum then $p'=a p_{\rm phys}$.
While $p'$ is conserved by the action,
the propagator is a matrix acting on the indices A.
As shown in Refs.  \cite{vandendoel,goltermansmit}, it can be written
(using the notation of Fig. \ref{feynmanfig})
\begin{eqnarray}
  G^{-1}(p,-q) &=&G^{-1}(p'+\pi A,-(q'+\pi B)) \\
  \label{propeqn2}
  &=&\bar\delta(q'-p') \left[
  m \iiden_{AB} - i \sum_\mu \sin(q'_\mu) {\ggam\mu}_{AB} \right] \ .
\end{eqnarray}
Here $\bar\delta(q'-p')$ is the periodic delta function, which sets
$q'_\mu=p'_\mu \ ({\rm mod}\ 2\pi)$, and there is an implicit
Kronecker delta for the color indices.

This way of writing the propagator makes the correspondence with a
four-flavor continuum theory obvious: the propagators have the same
form, except that different (but equivalent) bases are used.
The only significant change is the appearance of $\sin q'_\mu$
in Eq. \ref{propeqn2} rather than $q'_\mu$.
This means that lattice and continuum propagators differ by terms
of $O(a^2)$.

When using this propagator it is useful to keep two facts in mind.
First, the result actually holds for all $p'$ and $q'$,
i.e. one can lift the restriction that these momenta lie close to the origin.
Second, the matrices ${\ggam\mu}_{AB}$ simply permute (up to a sign)
the indices $A$ and $B$. Thus, for a given value of $A$,
the element of the matrix is
non-zero (and equal to $\pm1$) only for one value of $B$.

Using the results of Refs. \cite{goltermansmit,daniel}
the one gluon vertex can be written
\begin{equation}
  \label{vertexeqn}
  V_\mu(p'+\pi A,-(q'+\pi B), -k) = -i g\ T_c\ \bar\delta(q'+k-p')\
  \cos(q'_\mu +k_\mu/2)\ \ggam\mu_{AB} \ ,
\end{equation}
where $\mu$ is the gluon polarization direction, $c$ is its color,
and the momenta are defined in Fig. \ref{feynmanfig}.
Again this form is valid for any choices of $q'$ and $p'$,
and differs from the continuum vertex at $O(a^2)$.

We need the two gluon vertex only for tadpole diagrams.
In these, the total momentum of the two gluons entering the vertex is
zero, and both the gluons have the same color.
Thus the full vertex (given in Refs. \cite{goltermansmit,daniel})
simplifies to
\begin{equation}
  V_{\mu\nu} =  - i \half g^2 \ (T_c)^2\  \bar\delta(q'-p')\
  \delta_{\mu\nu}\ \sin(q'_\mu)\ {\ggam\mu}_{AB}  \ ,
\end{equation}
where $c$ is the color of the gluons, $\mu$ and $\nu$ their polarizations,
and the notation for momenta is as for the single gluon vertex.
This vertex, which is absent in the continuum,
is of $O(a^2)$, because there is, accompanying $g^2$,
an overall factor of $a$ (which we have set to 1)
which combines with the $a$ from the $\sin q'$.

Finally, the gluon propagator is
\begin{equation}
  D(k)_{\mu\nu} = \delta_{cc'}\ {1\over \sum_\sigma 4 \sin^2(k_\sigma/2)}
  \left[ \delta_{\mu\nu} - \alpha
  {4\sin(k_\mu/2)\sin(k_\nu/2) \over \sum_\rho 4 \sin^2(k_\rho/2)}
  \right] \ ,\\
\end{equation}
where $c$ and $c'$ are color indices, and $\mu$ and $\nu$ are polarizations.

Here $\alpha=0$ corresponds to Feynman gauge, $\alpha=1$ to Landau gauge.
The latter is defined on the lattice by maximizing
$\sum_{n,\mu} \Tr(U_\mu(n))$.
This may be ambiguous in general,
but gives a unique result in perturbation theory.

Ghost propagators are not required in our calculation.

\subsection{Lattice bilinears}

As mentioned in the introduction, we use a variety of lattice
transcriptions of the continuum operator $\co_{SF}^\CONT$.
These all make use of the ``hypercube fields'' introduced in
Ref. \cite{kluberg}.
To define these we divide the lattice in to $2^4$ hypercubes
in one of the 16 possible ways.
Points on the original lattice are then specified by a vector $y$
labeling the hypercubes (with all components even),
and a hypercube vector $C$ determining the position within the hypercube.
The 16 components of the staggered fermion field $\chi$ for a given $y$
are now collected in to a single hypercube field, which we label following
Ref. \cite{kieu}
\begin{equation}
  \chi(y)_C = \frac14 \chi(y+C) \ .
\end{equation}
In the continuum limit $\chi(y)_C$ becomes equal to the four-flavor
field $Q$, when expressed in the appropriate basis \cite{kluberg}
\begin{equation}
  \chi(y)_C \longrightarrow Q(y)_C = (\half\xi_C)^{\beta b} Q_{\beta,b} \ .
\end{equation}
Thus the lattice operator
\begin{equation}
\label{klubergeq}
  \co_{SF}(y) = \sum_{C,D}\ \chibar(y)_C\ \sf SFCD \ \chi(y)_D \ .
\end{equation}
has the same flavor, spin and normalization
as $\co_{SF}^\CONT=\bar Q \opergx SF Q$ in the continuum limit.
We often refer to $\co_{SF}$ by the abbreviated form $\chibar\sfno SF\chi$.
We also use $\sfno SF$ to denote $\co_{SF}$,
when confusion with the matrix that this notation also defines is unlikely.

\subsubsection{Gauge invariant bilinears}

In general, the operators defined by Eq. \ref{klubergeq} are gauge dependent,
since they have quark and antiquark fields at different sites.
We consider various ways of making the operators invariant
under gauge transformations.
First, we use the standard prescription of inserting the average of
the products of gauge links along the shortest paths
connecting the quark and antiquark.
These operators we refer to as ``gauge invariant'', or GI for short.

\begin{figure}
\vspace{1.truein}
\includegraphics{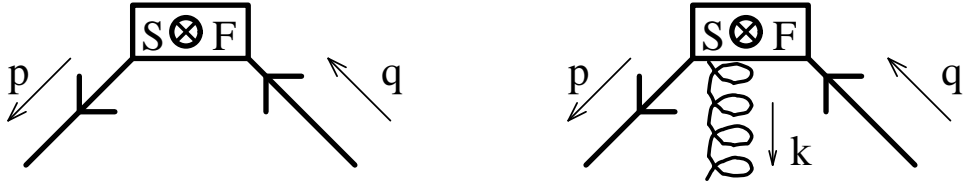}
\bigskip
\caption[operatorfig]{\fc
Notation for matrix elements of bilinears.}
\label{operatorfig}
\end{figure}

We need the vertex for an insertion of these operators between
an external quark and antiquark, and either zero, one or two gluons.
The insertion without gluons is (DS Eq. 14)
\begin{eqnarray}
  M_{SF}^{(0)}(p,-q) &=&{1\over 16} \sum_{CD} e^{i p\cdot C}\
  \sf SFCD \ e^{-i q\cdot D}  \\
  &=&\sum_{CD} e^{i p'\cdot C}\ {\frac14 (-)^{A\cdot C}}\
  \sf SFCD \ {\frac14 (-)^{D\cdot B}}\ e^{-i q'\cdot D} \ .
\end{eqnarray}
Here we use the momenta as defined in Fig. \ref{operatorfig},
and have set $y=0$ for convenience.
The fact that the operator is at a fixed position means that there
is no overall momentum delta function.
There is an implicit trace over color indices.

For this vertex to be useful in Feynman diagrams,
we must write it in terms of matrices in the ``momentum'' basis,
for this is the basis in which the flavor symmetry of the
propagators is apparent.
In fact, for physical momenta ($q',p'\to 0$)
the insertion becomes $\ssf SFAB$,
and is already in the momentum basis.
This shows that the operator is correct at tree level in the
continuum limit.
In loop diagrams, however, $q'$ and $p'$ do not vanish, and the
operator cannot be written in a simple form.
To make progress, DS use the following result
\begin{equation}
  e^{ik\cdot C} = \left( \prod_\mu e^{ik_\mu/2} \right)
  \sum_M E_M(k)\ (-)^{C\cdot \widetilde{M}}
\end{equation}
where
\begin{equation}
  \label{EMeqn}
  E_M(k) = \prod_\mu \half \left(
  e^{-ik_\mu/2} + (-)^{\widetilde{M}_\mu} e^{ik_\mu/2} \right) \ .
\end{equation}
Combining this with the Eq. \ref{gettildeeqn} one can rewrite the
insertion as
\begin{equation}
\label{lgeqn}
  M_{SF}^{(0)}(p'+\pi A,-(q'+\pi B))
  = \left( \prod_\mu e^{i(p'_\mu-q'_\mu)/2} \right)
  \sum_{MN} E_M(p')\ E_N(-q')\ \ssf{MSN}{MFN}AB \ ,
\end{equation}
which is the form we use for calculations.
Thus the original bilinear with spin $S$ and flavor $F$ has,
when inserted in loop diagrams, components with all possible
spins and flavors.

The single gluon vertex (see Fig. \ref{operatorfig} for notation)
has also been given by DS.
We find it convenient to slightly rewrite their result as
\begin{equation}
M_{SF}^{(1)\mu} (p,-q,k) = {i g T_c \over 16} \sum_{CD} (D\!-\!C)_\mu
  e^{ip\cdot C} e^{-iq\cdot D} e^{{i\over2}k\cdot (D\!+\!C)} \sf SFCD \
  h^\mu_{CD}(k) \ e^{-{i\over2} \sum_{\nu\ne\mu} k_\nu(D\!-\!C)_\nu} \ ,
\end{equation}
where the function $h^\mu_{CD}$ is related to the function
$f^\mu_{(CD)}$ of DS by
\begin{eqnarray}
  h^\mu_{CD}(k) &=& f^\mu_{(CD)}(k)
  \exp[-\half ik\cdot(D\!+\!C)] \
  \exp[\half i\sum_{\nu\ne\mu} k_\nu(D\!-C\!)_\nu] \\
  &=&{1\over12} \sum_{\nu\ne\mu} \left[
  1 + e^{ik_\nu(D\!-\!C)_\nu} + e^{i[k\cdot(D\!-\!C)
    - k_\mu(D\!-\!C)_\mu]} +    e^{i[k\cdot(D\!-\!C)
    - k_\mu(D\!-\!C)_\mu - k_\nu(D\!-\!C)_\nu]} \right] .
\end{eqnarray}
There are no implicit sums in this formula.
This form of the vertex is useful because the function
$h^\mu_{CD}(k)$ is independent of $k_\mu$.

The two gluon vertex is, in general, rather complicated.
We need it, however, only for tadpole diagrams,
and for these there is a considerable simplification.
We postpone our discussion of the two gluon vertex
until we calculate the tadpole diagrams.

\subsubsection{Projected bilinears}

One problem with the operators just discussed is that their matrix elements
are suppressed due to fluctuations of the intervening gauge links.
Much of this suppression comes from the tadpole diagrams.
The quadratic divergence of these diagrams, when combined with powers of
lattice spacing coming from expansion of $U_\mu$,
leave corrections which vanish in the continuum limit
as powers of $g^2$ (and not by powers of lattice spacing).
A solution to this problem was suggested by Sheard \cite{sheardtad},
who summed up the infinite series of tadpole graphs.
Another solution, proposed by Lepage and Mackenzie \cite{lepagemackenzie},
is to use the mean value of the intervening gauge links as a
renormalization factor for the bilinears.
We have considered still another option \cite{wius},
namely to project the averaged path ordered product of
gauge links back to the gauge group manifold.
We refer to the resulting operators as ``projected bilinears''.
As discussed further below, the projection amounts to the replacement
\begin{equation}
  {1 \over N_{path}} \sum_{path}\ \cp\ \exp [ig \int A_\mu dx^\mu ]
  \longrightarrow
  \exp [ig {1 \over N_{path}} \sum_{path}\ \int A_\mu dx^\mu + O(g^2)] \ ,
\end{equation}
where the ``$O(g^2)$'' term makes no contribution in the present calculation.
This replacement only affects bilinear operators containing
more than one gauge link. Moreover, it only alters the terms quadratic
or higher order in $A_\mu$, so that only tadpole diagrams are affected.
We discuss the changes in the vertices when we calculate these diagrams.

\subsubsection{Landau gauge bilinears}

We can avoid making the lattice operators explicitly
gauge invariant by choosing a specific gauge
(modulo ambiguities due to Gribov copies when they exist).
The simplest method is to fix to Landau gauge\footnote{%
This brings the gauge links as close to identity as possible.},
and then use the operator given in Eq. \ref{klubergeq}
without any gauge links.
There is an implicit trace of the color indices of the $\chibar$ and
$\chi$ fields.
The resulting operators, which we refer to as
``Landau gauge'' bilinears, have the same zero-gluon vertex
as the gauge invariant operators (Eq. \ref{lgeqn}),
but vertices involving gluons vanish.

\subsubsection{Smeared bilinears}

All the bilinears discussed so far
differ from the continuum bilinears at $O(a)$.
To see this consider the result for the zero-gluon vertex, Eq. \ref{lgeqn}.
The $O(a)$ terms come from the factors of $E_M$ and $E_N$ \footnote{%
It might appear the the first factor on the r.h.s. of Eq. \ref{lgeqn}
would give $O(a)$ differences between lattice and continuum operators.
This is not so. This factor comes from the fact that the lattice operator
should be associated with a continuum operator located at the center
of the hypercube, rather than at one of the corners as we have chosen.
This difference is irrelevant for the calculations we do here,
in which $p'=q'$ always.}.
It is easy to see from the definition, Eq. \ref{EMeqn}, that
$E_M(p')\sim a^{|\widetilde M|^2}$, if $p'\sim O(a)$.
Since $|\widetilde M|^2$ only vanishes when $M=(0,0,0,0)$,
it follows that all flavor breaking is suppressed by powers of $a$,
as we saw explicitly above.
The leading corrections occur for $|\widetilde M|=1$ and $|N|=0$,
or for $|M|=0$ and $|\widetilde N|=1$,
and are suppressed by only one power of $a$.

For operators containing gauge links there are additional $O(a)$ terms,
coming from the fact that the gauge fields are at varying positions
in the hypercube.
For Landau gauge bilinears, however, there are no such additional
corrections, and thus it is relatively straightforward to ``improve''
the operators so that they have, at tree level, only $O(a^2)$ corrections.

The source of the $O(a)$ corrections is the
non-locality of the staggered operators.
The quark and antiquark fields have slightly
different phases since they are at different points in the hypercube.
To remove the corrections one must smear the fields
so that their average position is the center of the hypercube.
One method for doing this was suggested in Ref. \cite{book}:
take the standard bilinear of Eq. \ref{klubergeq},
replace the quark field according to
\begin{equation}
  \chi(y)_A \to \frac14 \sum_\mu \chi(y+ 2\hat\mu[1-2A_\mu] )_A =
           \frac1{16} \sum_\mu \chi(y+A+ 2\hat\mu[1-2A_\mu] ) \ ,
\end{equation}
and perform a similar replacement for the antiquark field.
The resulting operators are spread out over a $4^4$ hypercube.
The Landau gauge version of these operators,
which are the only ones that we use,
then have only $O(a^2)$ corrections in tree level matrix elements.
We refer to them as ``smeared'' operators.

The Feynman rule for an insertion of a smeared operator
has the same form as that without smearing, except that
the functions $E_M$ are changed, i.e.
\begin{equation}
  \label{smeqn}
  M_{SF}^\SM(p'+\pi A,-q'-\pi B)
  = \left( \prod_\mu e^{i(p'_\mu-q'_\mu)/2} \right)
  \sum_{MN} E_M^\SM(p')\ E_N^\SM(-q')\ \ssf{MSN}{MFN}AB \ ,
\end{equation}
where
\begin{equation}
\label{EMSMeqn}
  E_M^\SM(k) =  E_M(k)\ \frac14 \sum_\mu
  \left({ e^{3ik_\mu/2} + (-)^{\widetilde M_\mu} e^{-3ik_\mu/2}
  \over e^{-ik_\mu/2} + (-)^{\widetilde M_\mu} e^{ik_\mu/2} }\right) \ .
\end{equation}
Both $E_M$ and $E_M^\SM$ share the important symmetry property
\begin{equation}
  k_\mu \to -k_\mu: \quad E_M(k) \to (-)^{\widetilde M_\mu} E_M(k)\ ,\
                      E_M^\SM(k) \to (-)^{\widetilde M_\mu} E_M^\SM(k) \ .
\end{equation}
This turns out to guarantee that the operator mixing is the same
for smeared and unsmeared operators.
It is simple to see using Eq. \ref{EMSMeqn} that, for $k\sim O(a)$,
the corrections are all of $O(a^2)$
\begin{equation}
  E_M^\SM(k) =  \left\{
  \begin{array}{ll} 1 + O(a^2) 	   &: |\tM|=0 \\
                    O(a^2)         &: |\tM|=1 \\
                    O(a^{|\tM|^2}) &: |\tM|>1 \\
  \end{array} \right. \ ,
\end{equation}
which demonstrates that the smeared operators are indeed
improved at tree level.

\section{CALCULATIONAL DETAILS}
\label{sdetails}

In this section we present an outline of the calculation of the
one-loop lattice corrections to the various bilinears defined above.
Those interested only in the results can skip to Section \ref{sresults}.
For the gauge invariant operators, the calculation has been done by DS,
and we restate only those portions of that work which are
needed to extend the calculation to other operators.
We also show some simplifications in the results of DS that we have found.

\begin{figure}
\vspace{3.truein}
\includegraphics{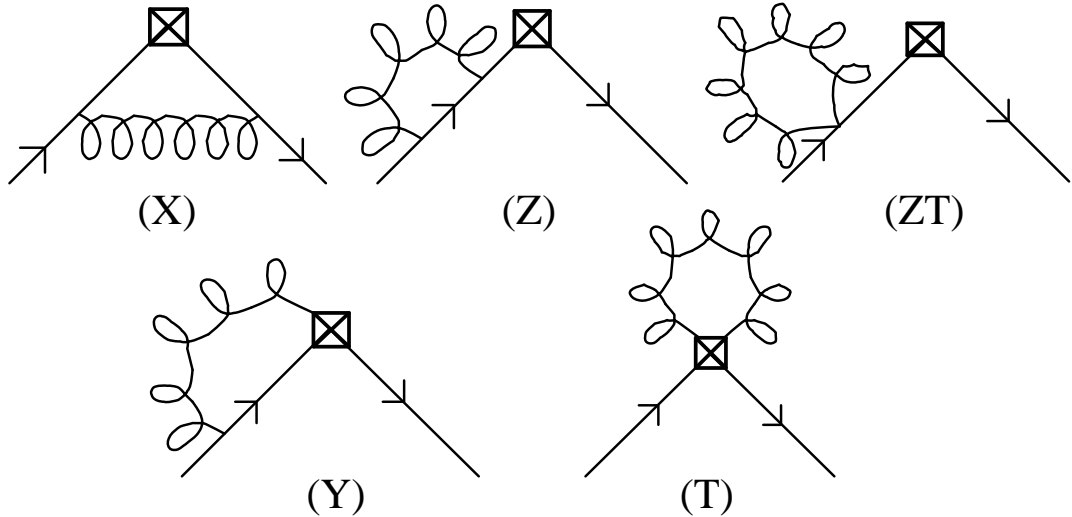}
\bigskip
\caption[oneloop]{\fc
Diagrams contributing to the renormalization of fermion bilinears.}
\label{oneloop}
\end{figure}

The diagrams which contribute are shown in Fig. \ref{oneloop},
where they are labeled following the notation of DS.
The ``X'', ``Z'' and ``ZT'' diagrams are common to all bilinears,
while the ``Y'' and ``T'' (for tadpole) graphs
are absent for Landau gauge operators.
For calculations involving Landau gauge operators
one must use the lattice Landau gauge gluon propagator,
while the choice of gauge is irrelevant for the gauge invariant operators.

Our aim is to determine the relationship between continuum and lattice
operators. To do this, we compare their matrix elements with $q\bar q$
external states. We set the external {\em physical} momenta to zero,
so that the components of the external {\em lattice} momenta are
either $0$ or $\pi$.
We label the momentum of the outgoing and incoming quarks
as $p=A\pi$ and $q=B\pi$ respectively.
We set the quark mass to zero,
since terms proportional to $m$ are suppressed by powers of $a$.

At tree level, all lattice versions of the continuum operator $\co_{SF}$
have been constructed to have the same matrix element,
\begin{equation}
\label{treelattice}
  \cm_{SF}^0(\LATT) = \ssf SFAB.
\end{equation}
The superscript indicates the number of loops.
At one-loop, the matrix elements of the various operators differ.
Each can be written
\begin{eqnarray}
  \cm_{i}^1(\LATT) &=& C_F\ {g^2 \over 16\pi^2}\ \cmb_i \\
  \label{latttree}
  \cmb_i  &=& d_i\ I_{div}\ \cm_i^0(\LATT)
                         + c_{ij}^1\ \cm_j^0(\LATT) \ ,
\end{eqnarray}
where the labels $i$ and $j$ each run over the 256 spin-flavor combinations.
The coefficients $d_i I_{div}$ give the divergent parts of the corrections,
and are the same for all versions of the lattice operator.
They are defined precisely below.
The coefficients $c_{ij}^1$ give the remaining finite corrections,
which do depend on the type of lattice operator.
The color factor, $C_F=(N^2-1)/2N$ for the gauge group $SU(N)$,
is ubiquitous and so we have factored it out, along with $g^2/(16\pi^2)$,
to define the reduced matrix element $\cmb_i$.

The X- and Z-diagrams are infrared divergent;
to regulate them we use a gluon mass $\lambda$,
the dependence on which cancels
when we compare lattice and continuum matrix elements.
Following GS, we isolate the divergent parts using the integral\footnote{%
DS use a slightly different regularization, in which the gluon mass
appears in both propagators. The result differs by a constant which
cancels in the comparison between lattice and continuum operators.}
\begin{equation}
  I_{div} = \int_\phi\
  {1\over (4 \sum_\mu \sbar_\mu^2)}\
  {1\over (4 \sum_\mu \sbar_\mu^2 + \lambda^2a^2)} \ .
\end{equation}
where we use the definitions
\begin{equation}
\label{gluonmass}
  \int_\phi \equiv 16 \pi^2 \int_{-\pi}^{\pi} {d^4\phi \over (2\pi)^4} \ ,
  \quad \sbar_\mu = \sin (\phi_\mu/2) \ ,
  \quad \cbar_\mu = \cos (\phi_\mu/2) \ .
\end{equation}
The result for this integral is given in GS:
\begin{equation}
  I_{div} =  -\ln(\lambda^2a^2) + F_{0000} - \gamma_E + 1 \ , \quad
  F_{0000}=4.36923(1) \ , \quad \gamma_E=0.577216... \ .
\end{equation}
The coefficients of $I_{div}$ are labeled $d_i$ in Eq. \ref{latttree},
and are proportional to the anomalous dimensions of the operators.
These corrections do not cause mixing amongst the bilinears,
as is well known in the continuum.

\begin{table}
\begin{center}
\begin{tabular}{cccccc}
Type
&$X_{ij}$        &$Y_i$     &$Z$       &$ZT$        &$T_i$     \\
\hline
GI (Feynman)
&X(F) 	         &Y(F)      &Z(F)      &ZT(F)       &T(F)      \\
GI (Landau)
&X(F) +X(L)      &Y(F)+Y(L) &Z(F)+Z(L) &ZT(F)+ZT(L) &T(F)+T(L) \\
Unsmeared L
&X(F) +X(L)	 &          &Z(F)+Z(L) &ZT(F)+ZT(L) & \\
Smeared L
&X(F,SM)+X(L,SM) & 	    &Z(F)+Z(L) &ZT(F)+ZT(L) & \\
Projected
&X(F) 		 &Y(F)      &Z(F)      &ZT(F)       &T(F,PR)   \\
\end{tabular}
\end{center}
\caption[coeffstab]{\fc
Labels for the contributions of the various diagrams
to the different types of lattice operator.
Corrections for gauge invariant (GI) operators
were calculated in DS using Feynman gauge (F).
Corrections for Landau gauge (L), smeared (SM) and projected (PR)
operators are calculated below.
Blank entries indicate that the diagrams do not contribute.}
\label{coeffstab}
\end{table}

All diagrams contribute to the finite coefficients $c_{ij}^1$,
and it is useful to decompose them as
\begin{equation}
  c_{ij}^1 = X_{ij} + \delta_{ij} (Y_i + Z + ZT + T_i) \ .
\end{equation}
This equation incorporates the result that only the X-diagrams cause mixing.
The fact that the Y-diagrams do not cause mixing is not apparent from the
results presented in DS, although it does follow from their numerical results.
We present below an explicit demonstration of this fact.
Furthermore, we show that $Y_i$ depends
only on the number of links in the bilinear, $\Delta = |S-F|^2$.
This is also true of the tadpole contribution, $T_i$.
Wave function renormalization is spin-flavor independent,
so that Z and ZT do not depend on $i$.
The coefficients X, Z, ZT, Y and T depend on which type of lattice
operator we use, as summarized in Table \ref{coeffstab}.
DS have calculated these coefficients for the gauge invariant operators
in Feynman gauge
(i.e. the coefficients in the top row of Table \ref{coeffstab});
we fill in the remainder of the table in the following.

\subsection{X-diagrams (Feynman gauge)}

The calculation of $X(\FSM)$, the Feynman gauge part of the X-diagram
contribution for smeared operators, follows precisely the same steps
as presented in DS for the unsmeared operators.
There is a divergent part depending only on the spin
\begin{equation}
	\label{sigmaSeqn}
  d_{SF}(X) = \sigma_S \ ,\qquad
  \frac14 \sum_{\mu\nu} \g\mu \g\nu \g{S} \g\nu \g\mu = \sigma_S\ \g{S} \ ,
  \end{equation}
and a finite part
\begin{eqnarray}
  \cmb^{X(\FSM)}_{SF}
  &=& \sum_{\mu\rho\sigma MN} X(\FSM)_{M}^{\mu,\rho\sigma}\
      \delta_{M,N+\hat\rho+\hat\sigma}\
      \ssf{\mu\rho MSN\sigma\mu}{MFN}AB \\
  &=& \sum_{\mu\rho\sigma M}  X(\FSM)_{M}^{\mu,\rho\sigma}\
      (-)^{\widetilde{M}\cdot(S+F) + S_\mu+S_\rho}\
      \zeta_\rho(F) \zeta_\sigma(F\!+\!\rho)
      \ssf{S}{[F+\rho+\sigma]}AB .
\end{eqnarray}
The coefficients $c_{ij}^1$ can be read off from this equation.
There is mixing between different flavors, but the spin remains unchanged.
The integrals in this expression are
\begin{equation}
  X(\FSM)_{M}^{\mu,\rho\sigma} = \int_\phi \left[
  \cbar_\mu^2 s_\rho s_\sigma B F^2
  \ E_M^\SM(\phi) \ E_{M+\hat\rho+\hat\sigma}^\SM(-\phi)
  - \frac14 \delta_{\rho\sigma} \delta_{M,{0}} B^2 \right] \ ,
\end{equation}
where
\begin{equation}
  s_\mu \equiv \sin(\phi_\mu) \ ,\quad
  c_\mu \equiv \cos(\phi_\mu) \ ,\quad
  F = (\sum_\nu s_\nu^2)^{-1} \ ,\quad
  B = (4 \sum_\nu \sbar_\nu^2)^{-1} \ .
\end{equation}
The symbols ``B'' and ``F'' here stand for the functions coming from
boson and fermion propagators respectively.
The expressions for the unsmeared operators are identical
except that $E_M^\SM\to E_M$.

\begin{table}
\begin{center}
\begin{tabular}{cccrr}
    M  & $X(\FSM)_M^{1,11}$ & $X(\FSM)_M^{2,11}$
       & $X(\FSM)_M^{1,12}$ & $X(\FSM)_M^{3,12}$ \\ \hline
  0 0 0 0 & -0.34392 &-0.40583 & 0.08410 \quad & 0.11908 \quad \\
  1 0 0 0 & -0.06584 &-0.01708 &-0.02214 \quad &-0.00446 \quad \\
  0 1 0 0 & -0.03842 &-0.09339 &-0.02214 \quad &-0.00446 \quad \\
  0 0 1 0 & -0.03842 &-0.01981 & 0.03534 \quad & 0.05691 \quad \\
  0 0 0 1 & -0.03842 &-0.01981 & 0.03534 \quad & 0.00878 \quad \\
  1 1 0 0 & -0.05870 &-0.03294 & 0.08410 \quad & 0.11908 \quad \\
  1 0 1 0 & -0.05870 &-0.13458 &-0.03323 \quad &-0.00820 \quad \\
  1 0 0 1 & -0.05870 &-0.13458 &-0.03323 \quad &-0.05550 \quad \\
  0 1 1 0 & -0.08962 &-0.02584 &-0.03323 \quad &-0.00820 \quad \\
  0 1 0 1 & -0.08962 &-0.02584 &-0.03323 \quad &-0.05550 \quad \\
  0 0 1 1 & -0.08962 &-0.11053 & 0.02303 \quad & 0.00461 \quad \\
  0 1 1 1 & -0.13870 &-0.26084 &-0.07130 \quad &-0.10743 \quad \\
  1 0 1 1 & -0.14808 &-0.05791 &-0.07130 \quad &-0.10743 \quad \\
  1 1 0 1 & -0.14808 &-0.17988 & 0.03534 \quad & 0.00878 \quad \\
  1 1 1 0 & -0.14808 &-0.17988 & 0.03534 \quad & 0.05691 \quad \\
  1 1 1 1 & -0.02866 &-0.01419 & 0.02303 \quad & 0.00461 \quad \\
\end{tabular}
\end{center}
\caption[xsmint]{\fc
X integrals for smeared operators, $X(\FSM)_M^{\mu,\rho\sigma}$.
Results are accurate to $\pm 0.00001$.}
\label{xsmint}
\end{table}

We have calculated these integrals both for smeared and unsmeared
operators. The results for the latter agree with those given in DS.
The former are collected in Table \ref{xsmint}.
The components not shown can be reconstructed using the symmetry
of the integral under $\rho\sigma$ interchange,
and by permuting the indices.

\subsection{X-diagrams (Landau gauge)}

The contribution of the Landau gauge part of the gluon propagator
to the X-diagrams is straightforward to evaluate.
The reason is that, as in the continuum, the
contraction of the $k^\mu$ in the gluon propagator with the $\gamma_\mu$ in
the vertex cancels with the adjacent fermion propagator, in the limits
of vanishing fermion mass and vanishing external momenta.
Thus the fermion propagators disappear, and the diagram reduces to a tadpole.
It turns out to be diagonal in spin-flavor space.

We now demonstrate this in more detail for the unsmeared operators.
The substitution $E_M\to E_M^\SM$ converts the calculation to
that for the smeared operators.
Using the Feynman rules given above, we find the following
contribution to the reduced matrix element
\begin{equation}
  \cmb^{X(L)}_{SF} =
  \sum_{\mu\nu\rho\sigma M N}
  \ssf{\rho\mu MSN \nu\sigma}{MFN}AB
  \int_\phi B F^2 \cbar_\rho s_\mu E_M(\phi) E_N(-\phi)
  s_\nu \cbar_\sigma (-4\sbar_\rho \sbar_\sigma B) \ .
\end{equation}
Combining terms using $s_\mu = 2 \sbar_\mu \cbar_\mu$ we see that
the integral is symmetric under the interchanges $\rho\leftrightarrow\mu$
and $\nu\leftrightarrow\sigma$. Thus we can make the replacements
$\ggam{\mu\rho} \rightarrow \delta_{\mu\rho} \iiden$, and
$\ggam{\nu\sigma} \rightarrow \delta_{\nu\sigma} \iiden$.
The expression then simplifies to
\begin{equation}
  \cmb^{X(L)}_{SF} = - \sum_{M N} \ssf{MSN}{MFN}AB
  \int_\phi B^2 E_M(\phi) E_N(-\phi) \ .
\end{equation}
As promised, the fermion propagators have been canceled.
Next we notice that the integral vanishes by symmetry unless $M=N$,
and so we find the final expression
\begin{equation}
  \cmb^{X(L)}_{SF} = - \ssf{S}{F}AB \sum_{M} (-)^{\widetilde{M}\cdot(S-F)}
  \int_\phi B^2 E_M(\phi) E_M(-\phi) \ .
\end{equation}
This shows that the correction is diagonal in spin and flavor.

The integral is divergent for $M=0$, and gives a contribution
to $d_i$ of $d(X(L))=-1$ independent of spin or flavor,
for both smeared and unsmeared operators.
The finite parts are
\begin{equation}
\label{xlandresult}
  X(L)_{SF}= \sum_M (-)^{\widetilde{M}\cdot(S-F)} \int_\phi
  \left( \delta_{M0} B^2 - B^2 E_M(\phi) E_M(-\phi) \right) \ .
\end{equation}
It is simple to see from the definition of $E_M$
that the integral depends only on $|\tM|^2$.
It then follows that $X(L)_{SF}$ depends only on $\Delta=|S-F|^2$:
$X(L)_{SF} = X(L)_{\Delta}$.
The numerical values for these quantities,
for both smeared and unsmeared operators,
are given in Table \ref{xlints}.

For the unsmeared operators, it is possible to derive a simpler
form for the result. This can be done, for example,
by inserting the definition of $E_M$ into Eq. \ref{xlandresult}.
The results are
\begin{equation}
  \label{xldelta}
  X(L)_{\Delta=0} = 0 \ ,\
  X(L)_{\Delta\ge1} = \int_\phi B^2 (1 - \Pi_{i\le\Delta} c_i) \ .
\end{equation}

\begin{table}
\begin{center}
\begin{tabular}{crr}
   $\Delta$  &$X(L)_\Delta$ \quad &$X(L,\SM)_\Delta$ \\ \hline
	 0   & 0.00000     & 4.19152  \\
	 1   & 3.05826     & 4.97458  \\
	 2   & 4.07769     & 5.12452  \\
	 3   & 4.59516     & 5.17390  \\
	 4   & 4.92625     & 5.19848  \\
\end{tabular}
\end{center}
\caption[xlints]{\fc
Landau gauge part of X integrals for unsmeared and
smeared operators. The results depend only on $\Delta=|S-F|^2$.
They are accurate to $\pm 0.00001$.}
\label{xlints}
\end{table}

\subsection{Y-diagrams (Feynman gauge)}

We review here the calculation of Y-diagrams---those in which
a gluon joins an external fermion leg to a link in the operator.
Clearly these diagrams are absent for Landau gauge operators.

Consider the matrix element of the bilinear $\sfno SF$.
Using the Feynman rules given above,
standard manipulations yield
(for this calculation it is simpler to display the result using
the ``single-bar'' spin-flavor representation)
\begin{eqnarray}
\nonumber
 \cmb^{Y(F)}_{SF} &=&
 - i \sum_{CDE} \sum_{\mu\rho} I^Y(\rho,\mu,D\!-\!C) \times
    \left[ {\frac14 (-)^{A\cdot E}} \gam{\mu\rho}_{EC} \sf S F C D
           {\frac14 (-)^{D\cdot B}} \right. \\
\label{ydiagres}
&& \qquad\left. \mbox{} +
{\frac14 (-)^{A\cdot C}} \sf S F C D \gam{\rho\mu}_{DE}
{\frac14 (-)^{E\cdot B}} \right] \ ,
\end{eqnarray}
where the integral is
\begin{equation}
  I^Y(\rho,\mu,D\!-\!C) = (D\!-\!C)_\mu \int_\phi B F \ s_\rho \ \cbar_\mu \
                      h^\mu_{CD}(\phi) \exp[\half i\phi_\mu(D\!-\!C)_\mu] \ .
\end{equation}
The first term in Eq. \ref{ydiagres} comes from
the gluon attaching to the outgoing quark line,
the second from attachment to the incoming quark.

Expanding the exponential gives a sine and a cosine term
\begin{equation}
\label{expexpansion}
\exp[\half i\phi_\mu (D\!-\!C)_\mu] = i (D\!-\!C)_\mu \sbar_\mu +
\cbar_\mu \ .
\end{equation}
We show below that the cosine term does not contribute,
so we keep only the sine term.
Since the function $h^\mu$ does not depend on the $\mu$-th
component of $\phi$, the integral over $\phi_\mu$ vanishes unless $\rho=\mu$.
Thus we can sum over $E$ giving
\begin{equation}
\label{ydiagres2}
 \cmb^{Y(F)}_{SF} = \sum_{CD}
  {(-)^{A\cdot C}\over4} \sf S F C D {(-)^{D\cdot B}\over4}
  \sum_\mu (D\!-\!C)_\mu^2 \int_\phi B F \ s_\mu^2 \ h^\mu_{CD}(\phi) \ .
\end{equation}
{}From the definition of $h^\mu$, it is simple to see that
the integral depends on the spin-flavor of the operator
only through the combination $\Delta = |D\!-\!C|^2$,
the number of gauge links.
This allows us to sum over $C$ and $D$,
and see explicitly that there is no mixing between different spin-flavors
\begin{equation}
\label{ydiagres3}
 \cmb^{Y(F)}_{SF} = \ssf S F A B Y(F)_{\Delta} \ .
\end{equation}
The results are conveniently expressed as
\begin{equation}
  Y(F)_\Delta = Y(F)_{\Delta-1} + I_\Delta \ ,
\end{equation}
where the integrals are
\begin{eqnarray}
  I_1 &=& \int_\phi B F s_1^2 = \int_\phi B / 4 \ , \\
  I_2 &=& \int_\phi B F s_1^2 c_2         \ , \\
  I_3 &=& \int_\phi B F s_1^2 c_2 c_3     \ , \\
  I_4 &=& \int_\phi B F s_1^2 c_2 c_3 c_4 \ .
\end{eqnarray}
Of course, $Y(F)_{\Delta=0}=0$,
for there is no Y-diagram if the operator has no gauge links.
We tabulate the numerical values for these constants in Table \ref{yints}.
We have checked the results numerically against those obtained by
appropriately summing the terms in Table 3 of DS.\footnote{%
The signs of the constants, labeled $Y_{MN}^{1,1}$ by DS, were
inadvertently left out from their Table 3; our results correct this error.}

\begin{table}
\begin{center}
\begin{tabular}{ccc}
   $\Delta$  &$Y(F)_\Delta$ &$Y(L)_\Delta$ \\
\hline 
        0    & 0.00000   & -0.00000      \\
	1    & 6.11653   & -6.11653      \\
	2    & 7.42832   & -8.15537      \\
	3    & 8.02785   & -9.19031      \\
	4    & 8.41043   & -9.85250      \\
\end{tabular}
\end{center}
\caption[yints]{\fc
Y-integrals in Feynman gauge, and from the Landau gauge part of
the gluon propagators.
The results depend only on $\Delta=|S-F|^2$,
and are accurate to $\pm 0.00001$.}
\label{yints}
\end{table}

Now we argue that the cosine term in Eq. \ref{expexpansion}
gives a null result.
This is true separately for the diagrams in which the gluon attaches to
the incoming and outgoing quarks.
In each diagram, $s_\rho$ must pair with one of the $s_{\nu\ne\mu}$
coming from $h^\mu_{CD}$ to produce a non-zero result.
The resulting contribution to $I^Y(\rho,\mu,D\!-\!C)$ is
\begin{eqnarray}
  &{i\over12}& (D\!-\!C)_\mu (D\!-\!C)_\rho
    \int_\phi B(\phi) F(\phi) s_\rho^2 \ \cbar_\mu^2 \\
\nonumber
  && \times \left[
  1 + \cos((D\!-\!C)_\sigma\phi_\sigma) + \cos((D\!-\!C)_\tau\phi_\tau)
    + 3 \cos((D\!-\!C)_\sigma\phi_\sigma)
        \cos((D\!-\!C)_\tau\phi_\tau) \right]
\end{eqnarray}
with $\mu,\rho,\sigma,\tau$ all distinct.
The integral is just a number, depending on $\Delta$ but independent of
$\mu$ and $\rho$. Thus the contribution to $I^Y$ is a symmetric function
of $\mu$ and $\rho$, but with the restriction that $\rho\ne\mu$.
Such a function contracted with $\g{\mu\rho}$ obviously gives zero.
Thus the off-diagonal part of the Y-diagram vanishes.


\subsection{Y-diagrams (Landau gauge)}

The calculation of the Landau gauge part of the Y-diagrams turns out to be
very straightforward. The extra term in the Landau gauge gluon propagator
again combines with the vertex to exactly cancel the fermion propagator.
This leaves behind a tadpole graph with only diagonal spin-flavor corrections.
The contribution is the same for both possible attachments of the gluon.
The resulting renormalization is
\begin{equation}
 \cmb^{Y(L)}_{SF} = -4 \sum_{CD}
  {\frac14 (-)^{A\cdot C}} \sf S F C D {\frac14 (-)^{D\cdot B}}
  \sum_\mu (D\!-\!C)_\mu^2
  \int_\phi B^2\sbar_\mu^2 \ h^\mu_{CD}(\phi)  \ .
\end{equation}
The integral again depends only on the number of links
in the operator, $\Delta$, so that we can sum over $C$ and $D$.
Working out the cases individually,
the results take the same form as the Feynman gauge Y-diagram results,
\begin{equation}
  \label{yldelta}
  Y(L)_\Delta = Y(L)_{\Delta-1} + I^L_\Delta \ , \quad
  Y(L)_{\Delta=0} = 0 \ ,
\end{equation}
with the integrals given by
\begin{eqnarray}
  I^L_1 &=& - 4 \int_\phi B^2 \sbar_1^2 = - \int_\phi B / 4 \ , \\
  I^L_2 &=& - 4 \int_\phi B^2 \sbar_1^2 c_2         \ , \\
  I^L_3 &=& - 4 \int_\phi B^2 \sbar_1^2 c_2 c_3     \ , \\
  I^L_4 &=& - 4 \int_\phi B^2 \sbar_1^2 c_2 c_3 c_4 \ .
\end{eqnarray}
Numerical integration yields the results quoted in Table \ref{yints}.
It is straightforward to show that these results are related
to those from the X-diagrams in Landau gauge by
$Y(L)_\Delta = - 2 X(L)_\Delta$.

\subsection{Self-energy diagrams}

The Z- and ZT-diagrams give rise to wave-function renormalization.
The lattice symmetries are sufficient to ensure that this
renormalization does not break the flavor symmetry \cite{sharatchandra},
so that the contribution to matrix elements is independent of
spin and flavor.
DS calculated these diagrams implicitly using Ward Identities.
We have checked their result explicitly,
and extended the calculation to Landau gauge.

We need to calculate the renormalization diagrams with
non-zero external physical momentum for the quark, $p' \ne 0$,
and then pick the term proportional to $p'$ in a Taylor series
expansion of the resulting expression.
(Of course, the term independent of $p'$ vanishes for massless quarks.)
Considerable algebra leads to the following results for the
contribution of the Z-diagram in Feynman gauge:
\begin{eqnarray}
  d(Z(F)) &=& -1 \ , \\
  Z(F)    &=& - \int_\phi \left[ 2 B F c_1 (2 - \cbar_1^2) (1 - 2 F s_1^2)
                               + {B\over4} - B^2 \right] \ .
\end{eqnarray}
Numerically we find $Z(F) = -0.90135(1)$.

The Landau gauge Z-diagram leads to the simple results:
\begin{eqnarray}
  d(Z(L)) &=& 1 \ , \\
  Z(L)    &=& \int_\phi {B\over8} = 3.058263(1) \ .
\end{eqnarray}
The divergent contribution satisfies two necessary conditions.
First, since the total divergent contribution from the Landau gauge
part of the gluon propagator must vanish, and since only X- and Z-diagrams
are divergent, it must be that $d(Z(L))+d(X(L))=0$.
Second, as is well known in the continuum, the self-energy in Landau gauge
is finite, so that $d(Z)+d(Z(L))=0$.

The tadpole self-energy diagrams yield the finite contributions
\begin{equation}
  ZT(F) =   \int_\phi {B\over2} \ , \quad
  ZT(L) = - \int_\phi {B\over8} \ , \quad
\end{equation}
Notice that $Z(L) + ZT(L) = 0$,
so that the the Landau gauge part of the gluon propagator has
no net effect on the finite part of the self-energy diagrams.
The total finite part of self-energy correction is thus
\begin{equation}
  Z(F) + ZT(F) = 11.33170(1) \ .
\end{equation}

\subsection{Tadpole diagrams}

The final part of the calculation is that of the tadpole diagrams.
These are obviously absent for the Landau gauge operators.
For the gauge invariant operators,
it is here alone that the projection of the sum of products
of gauge link matrices back in to the gauge group has an effect.

First let us recall the result if no projection has been done.
Then for an operator with $\Delta$ links,
the tadpoles in Feynman gauge give
\begin{equation}
  T(F)_\Delta = - \Delta \int_\phi {B\over 2}
              = - \Delta \times 12.233053(1) \ .
\end{equation}
To complete our check that Landau and Feynman gauges give the same results
we need to calculate the Landau gauge contribution to the tadpoles.
The diagonal (i.e. $\mu=\nu$) contribution from the Landau gauge part
of the gluon propagator is just $-\fourth T(F)_\Delta$. In addition,
the Landau gauge part of the gluon propagator allows off-diagonal
contributions between links in different directions (i.e. $\mu\ne\nu$).
Explicit evaluation gives for the total
\begin{equation}
T(L)_\Delta = - \fourth T(F)_\Delta - 4 \int_\phi B^2 \times
\left\{
  \begin{array}{ll} 0                                   &: \Delta=0,1 \\
                    \sbar_1^2 \sbar_2^2                 &: \Delta=2   \\
                    \sbar_1^2 \sbar_2^2 (2+c_3)         &: \Delta=3   \\
                    \sbar_1^2 \sbar_2^2 (3+2c_3+c_3c_4) &: \Delta=4   \\
  \end{array} \right. \ .
\end{equation}
It is easily verified that $T(L)_\Delta = -\half Y(L)_\Delta$.
Since we also have $X(L)_\Delta = -\half Y(L)_\Delta$,
the sum of Landau gauge contributions to the X,
the Y and the tadpole diagrams gives zero.
We have already seen that the infinite parts of the Landau gauge
contributions cancel, and that the finite parts contributing
to the self-energy diagrams cancel.
Thus we have completed our check that the extra part of the Landau gauge
gluon propagator gives no contribution to gauge invariant operators.
This gives us confidence that our calculation of the extra contribution
to Landau gauge operators, X(L), is correct.

\begin{figure}
\vspace{1.truein}
\includegraphics{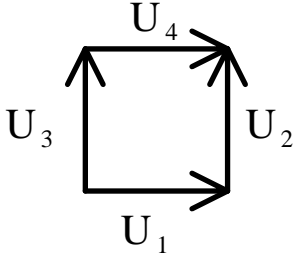}
\bigskip
\caption[twoproj]{\fc
2-link paths illustrating the effect of projection.}
\label{twoproj}
\end{figure}

Now we turn to the operators containing projected gauge links.
We only calculate the tadpole diagrams for these operators in Feynman gauge.
For 0- and 1-link operators projection has no effect.
We illustrate the calculation for 2-link operators.
Without the projection, the average of paths connecting the quark
and the antiquark is (see Fig. \ref{twoproj})
\begin{equation}
  \half ( U_1 U_2 + U_3 U_4 ) = 1 + {ig\over2} (A_1 + A_2 + A_3 + A_4)
  - {g^2\over4} (A_1^2 + A_2^2 + 2 A_1 A_2 + A_3^2 + A_4^2 + 2 A_3 A_4)
  + \cdots \ ,
\end{equation}
where the $A$'s are Hermitian traceless matrices.
In Feynman gauge only the diagonal parts of the $O(g^2)$ terms contribute,
and one gets the result quoted above, with $\Delta=2$.
The projection replaces this matrix with
\begin{eqnarray}
\nonumber
  1 & + & {ig\over2}  (A_1 + A_2 + A_3 + A_4)
    \ - \ {g^2\over4} ([A_1,A_2]+[A_3,A_4]) \\
    & - & {g^2\over8} (A_1 + A_2 + A_3 + A_4)\ (A_1 + A_2 + A_3 + A_4)
      + \cdots \ .
\end{eqnarray}
The $O(g)$ terms are unchanged, while the $O(g^2)$ terms are different.
For a pair of parallel links, say $U_1$ and $U_4$, the replacement is
\begin{equation}
  A_1^2 + A_4^2  \longrightarrow \half (A_1^2 + A_4^2 + A_1 A_4 + A_4 A_1) \ .
\end{equation}
The crucial point is that while $A_1^2$ gives the integral $\int_\phi B$,
cross terms like $A_1 A_4$ give $\int_\phi B c_1$ which is smaller.
Note that the $O(g^2)$ commutator terms in the projected
link do not contribute in Feynman gauge.


The tadpole contributions with projections are of course diagonal in
spin-flavor space, and depend only on the number of links in the bilinear.
We find
\begin{eqnarray}
  T(\FPR)_1 &=& -\half \int_\phi B \\
  T(\FPR)_2 &=& -\half \int_\phi B (1 + c_1) \\
  T(\FPR)_3 &=& -\half \int_\phi {B\over6} (5 + 8c_1 + 5c_1c_2) \\
  T(\FPR)_4 &=& -\half \int_\phi {2B\over3} (1 + 2c_1 + 2c_1c_2 + c_1c_2c_3)
     \ .
\end{eqnarray}
The numerical values are given in Table \ref{tadints}.
We are interested in the ratios by which the projection reduces
the size of the tadpoles, $T(\FPR)_\Delta/T_\Delta$.
We are also interested in the size of the projected tadpoles
relative to the single link tadpole, $T(\FPR)_\Delta/T(F)_1$.
Both of these quantities are also given in the Table.
The results show that the projection is largely successful in reducing
the size of the tadpole contribution to the level of that of a single link,
i.e. it reduces the penalty for introducing extra gauge links.

\begin{table}
\begin{center}
\begin{tabular}{cccc}
$\Delta$ &$T(\FPR)_\Delta$ &$T(\FPR)_\Delta/T(F)_\Delta$
                            &$T(\FPR)_\Delta/T(F)_1$ \\
\hline
 1    & -12.23305 & 1.00  & 1.00 \\
 2    & -14.59650 & 0.60  & 1.19 \\
 3    & -14.18207 & 0.39  & 1.16 \\
 4    & -13.05229 & 0.27  & 1.07 \\
\end{tabular}
\end{center}
\caption[tadints]{\fc
Tadpole integrals in Feynman gauge, with projection.
The results depend only on $\Delta=|S-F|^2$.
Results for $T(\FPR)_\Delta$ are accurate to $\pm 0.00001$.}
\label{tadints}
\end{table}

\section{RESULTS}
\label{sresults}

To relate lattice and continuum operators,
we must also do the one-loop calculation in the continuum.
Since the continuum bilinears are local,
only the X- and Z-diagrams contribute.
The tree level matrix elements are, by construction,
the same as those of the lattice operators
\begin{equation}
  \cm_{SF}^0(\CONT) = \opergx{S}{F} \ ,
\end{equation}
as long as the lattice results (Eq. \ref{treelattice})
are expressed in the appropriate basis.
The one-loop result is
\begin{eqnarray}
  \cm_{i}^1(\CONT) &=& C_F {g^2 \over 16\pi^2}
                       d_i^\CONT \cm_{i}^0(\CONT) \ , \\
  \label{conttree}
  d_i^\CONT &=& (\sigma_S-1)(-\ln(\lambda^2/\mu^2) + 1)  + t_S
\end{eqnarray}
where $\sigma_S$, defined in Eq. \ref{sigmaSeqn}, is $(4,1,0,1,4)$ for
spin tensors $(S,V,T,A,P)$,
$\lambda$ is the gluon mass which regulates infrared divergences
as in Eq. \ref{gluonmass},
$\mu$ is the renormalization scale,
and $t_S$ depends on the ultraviolet regulator.
We present results for four such regulators,
all of which have been used in previous calculations,
and which differ in their treatment of $\gamma_5$:
\begin{itemize}
\item[A.]
Dimensional reduction \cite{siegel} using the ``easy subtraction'' scheme of
Bernard, Draper and Soni \cite{uclapert} (\DREZ);
\item[B.]
Dimensional reduction using the subtraction scheme of
Altarelli {\em et al.} \cite{altarelli} (\DRED);
\item[C.]
Naive dimensional regularization with an anticommuting $\gamma_5$ (NDR);
\item[D.]
Dimensional regularization using the 't Hooft-Veltman
prescription for $\gamma_5$ \cite{thooft} (\HV).
\end{itemize}
All four regulators use modified minimal subtraction to remove the poles
at four dimensions.
The latter three schemes are discussed in detail in Ref. \cite{buras}.
We find
\begin{equation}
\label{tsequation}
t_S =
\left\{
\begin{array}{rrrrrrll}
(& 0.5,&  0,&0.5,&  0,& 0.5)&\qquad\qquad
				&(\DREZ) \\ 
(& 0.5,&0.5,&0.5,&0.5,& 0.5)&	&(\DRED) \\ %
(&-0.5,&  0,&1.5,&  0,&-0.5)&	&(\NDR)  \\ 
(&-0.5,&  0,&1.5,&  4,& 7.5)&	&(\HV)   \\ %
  \end{array} \right.
\end{equation}
for spin tensors $(S,V,T,A,P)$ respectively.

Combining Eqs. \ref{latttree} and \ref{conttree} we obtain the one-loop
operator matching relationships
\begin{equation}
  \label{contlatt}
  \co^{\CONT}_i = \co^{\LATT}_i  + C_F {g^2\over 16\pi^2}
  \sum_j \left(\delta_{ij} 2d_i \ln(\mu a/\pi) + c_{ij}\right) \co^{\LATT}_j
  \ ,
\end{equation}
where the $d_i = (\sigma_S-1)$ are the same as those appearing in
Eq. \ref{latttree}, and
\begin{equation}
  \label{mixingcs}
  c_{ij} = \delta_{ij} \left[
  (\sigma_S-1) (2\ln\pi -F_{0000} +\gamma_E) + t_S
                       \right] - c_{ij}^1 \ .
\end{equation}
The dependence on the infrared regulator has canceled,
leaving a finite logarithm.
We have chosen the scale of this log such that it vanishes if $\mu=\pi/a$,
which is the typical scale of momenta in lattice loop integrals.
In this way the coefficients $c_{ij}$ are a direct measure of the size
of the perturbative corrections.
We discuss this further in the next section.

\begin{table}
\begin{center}
\begin{tabular}{cccrrrr}
Operator		      &$d_i$&Name 	&
	(a) \quad &(b) \quad &(c) \quad &(d) \quad \\
\hline 
$\operii$		      & 3  &$c_{SS}$	&
        -31.3573 & -31.3573 & -31.3573 &  -8.6674 \\
$\operix 5$	      	      & 3  &$c_{SP}$  &
         46.2255 &  10.3456 &   0.7775 &   1.3880 \\
$\operix \mu$		      & 3  &$c_{SV}$  &
         -0.7743 &  -0.7743 &  -9.9491 &  -1.9304 \\
$\operix{\mu5}$		      & 3  &$c_{SA}$  &
         32.8444 &  10.3273 &  -0.4221 &   0.9857 \\
$\operix{\mu\nu}$	      & 3  &$c_{ST}$  &
         18.3027 &   8.4331 &  -2.8128 &   0.2204 \\
$\opergi{\mu}$		      & 0  &$c_{VS}$	&
          0.0000 &   0.0000 &  -9.1748 & -10.6286 \\
$\opergx\mu5$		      & 0  &$c_{VP}$	&
         22.5091 &  -0.0080 & -10.7574 & -11.1387 \\
$\opergx\mu\mu$		      & 0  &$c_{VV0}$	&
        -14.7798 & -14.7798 & -14.7798 & -12.6450 \\
$\opergx\mu\nu$		      & 0  &$c_{VV2}$	&
         10.0407 &   0.1711 & -11.0748 & -11.1052 \\
$\opergx\mu{\nu5}$	      & 0  &$c_{VA2}$	&
         10.0588 &   0.1892 & -11.0567 & -11.3259 \\
$\opergx\mu{\mu5}$	      & 0  &$c_{VA4}$	&
         34.5008 &  -1.3791 & -10.9472 & -11.0055 \\
$\opergx\mu{\mu\nu}$	      & 0  &$c_{VT1}$ &
         -3.3958 &  -3.3958 & -12.5706 & -11.8881 \\
$\opergx\mu{\nu\rho}$	      & 0  &$c_{VT3}$ &
         22.2422 &  -0.2749 & -11.0243 & -11.0590 \\
$\opergi{\mu\nu}$	      &-1  &$c_{TS}$  &
          8.1761 &  -1.6935 & -12.9394 & -14.3310 \\
$\opergx{\mu\nu}\mu$	      &-1  &$c_{TV1}$ &
         -2.4711 &  -2.4711 & -11.6459 & -14.1209 \\
$\opergx{\mu\nu}\rho$	      &-1  &$c_{TV3}$ &
         19.4637 &  -3.0533 & -13.8027 & -14.4338 \\
$\opergx{\mu\nu}{\mu\nu}$     &-1  &$c_{TT0}$ &
         -8.5873 &  -8.5873 &  -8.5873 & -13.3042 \\
$\opergx{\mu\nu}{\mu\rho}$    &-1  &$c_{TT2}$ &
          7.8601 &  -2.0095 & -13.2554 & -14.3758 \\
$\opergx{\mu\nu}{\rho\sigma}$ &-1  &$c_{TT4}$ &
         31.2592 &  -4.6207 & -14.1888 & -14.4700 \\
\end{tabular}
\end{center}
\caption[diagcoefs]{\fc
Results for the diagonal part of renormalization constants at one-loop,
i.e. $d_i$ and $c_{ii}$ using the \DREZ\ scheme in the continuum.
The components $\mu$, $\nu$, $\rho$ and $\sigma$ are all different.
The finite parts $c_{ii}$ are given for four choices of operators:
(a) gauge invariant;
(b) projected gauge links;
(c) unsmeared Landau gauge;
and (d) smeared Landau gauge.
Those renormalization constants not shown can
be obtained from these results, as explained in the text.
The error in the results is no larger than $0.0001$.}
\label{diagcoefs}
\end{table}

The diagonal matching coefficients are tabulated in Table \ref{diagcoefs},
for the \DREZ\ scheme.\footnote{%
The results quoted previously, in Refs. \cite{book,ssringberg},
have various errors which are corrected here.}
We give the results for only half of the operators,
because the mixing coefficients are the same for the two operators
$\opergx{S}{F}$ and $\opergx{S5}{F5}$.
Thus, for example, $c_{SS}=c_{PP}$.\footnote{%
For the operators with an even number of links this is due
to the axial symmetry \cite{smitvink}.
For operators with an odd number of links
we do not know of an argument why this equality should hold at higher orders.
The exception is the conserved vector and partially
conserved axial currents ($c_{VS}$ and $c_{AP}$),
for which the renormalization corrections should vanish.}
The names we give the coefficients indicate the spin and flavor
of the operator, and, if there is an ambiguity, the number
of links in operator, i.e. $\Delta=|S-F|^2$.
Thus $C_{VT1}$ corresponds to the vector current with ``tensor'' flavor
having only a single link.

The mixing coefficients are given in Table \ref{mixingcoefs}.
As noted by DS, the 1-loop diagrams do not cause mixing to the
full extent allowed by lattice symmetries \cite{verstegen}.
Since only the Feynman gauge part of the X-diagrams gives rise to mixing,
the off-diagonal coefficients are the same for gauge invariant,
Landau gauge and projected operators.
The only difference is between smeared and unsmeared operators.

\begin{table}
\begin{center}
\begin{tabular}{cccrr}
Operator-$i$ 		&Operator-$j$ 	       	&Name
					&(a) \quad     &(b) \quad \\
\hline 
$\opergx\mu\nu$	  	&$\opergx\mu\mu$	&$c_{VVM}$
					& 3.04128 	& 0.12152  \\
$\opergx\mu{\mu5}$	&$\opergx\mu{\nu5}$	&$c_{VAM}$
					&-0.64608  	& 0.20344  \\
$\opergx\mu{\mu\nu5}$	&$\opergx\mu{\rho\nu5}$	&$c_{VTM}$
					&-1.48592 	& 0.25360  \\
$\opergx{\mu\nu}{\mu5}$	&$\opergx{\mu\nu}{\rho5}$&$c_{TAM}$
					&-0.67632 	& 0.08712  \\
\end{tabular}
\end{center}
\caption[mixingcoefs]{\fc
Results for the off-diagonal part of renormalization constants at
one-loop, $c_{ij}$, for (a) unsmeared and (b) smeared operators.
The components $\mu$, $\nu$ and $\rho$ are all different,
and when $\rho$ appears it can take either of the two allowed values.
Mixing coefficients related by axial symmetry are not shown.
The error in the results does not exceed $0.00001$.}
\label{mixingcoefs}
\end{table}

To illustrate the use of these tables we give two examples.
A lattice operator which, in the continuum limit, matches onto the
vector current $\bar Q \opergx\mu\nu Q$ ($\mu\ne\nu$) is
\begin{equation}
  \chibar \sfno\mu\nu \chi + C_F {g^2\over 16\pi^2} \left[
  c_{VV2}\ \chibar \sfno\mu\nu \chi +
  c_{VVM}\ \chibar \sfno\mu\mu \chi \right] \ ,
\end{equation}
where there is no summation over any indices.
Our second example requires the use of the axial symmetry.
The lattice operator matching onto the axial current
$\bar Q \opergx{\mu5}{\mu\nu} Q$ is
\begin{equation}
  \chibar \sfno{\mu5}{\mu\nu} \chi + C_F {g^2\over 16\pi^2} \left[
  c_{VT3} \                            \chibar \sfno{\mu5}{\mu\nu} \chi +
  c_{VTM} \sum_{\rho\ne\mu,\rho\ne\nu} \chibar \sfno{\mu5}{\rho\nu} \chi
  \right] \ .
\end{equation}

As stated above, we have checked the results of DS for the gauge
invariant operators (Tables 4 and 5 of Ref. \cite{daniel}).
The relationship between our coefficients and those quoted by DS is
\begin{equation}
 c_{ij}^{\rm DS} = c_{ij}^1 + \delta_{ij}(\sigma_S-1)(F_{0000}- \gamma_E) \ .
\end{equation}
Using this, it is simple to compare the diagonal coefficients with those
of DS, although the off-diagonal coefficients require more work,
since DS use a different basis for the operators.

GS have also done the matching calculation for gauge invariant
operators of the form $\operix{F}$ for all flavors $F$.
They extract from their calculation finite flavor dependent constant,
$\sigma_F$ (which has no relationship to $\sigma_S$ defined above),
and a constant from wavefunction renormalization, $\tau$.
One can show analytically that these are related to our constants $c_{ij}$ by
\begin{equation}
  16\pi^2(\sigma_F-\tau) = c_{1\otimes F}^{\rm DS} - \half
                         = t_{S=1} + 6 \ln\pi - \half - c_{1\otimes F} \ .
\end{equation}
We have checked numerically that these relationships hold.

\section{DISCUSSION}
\label{sconcl}

How large, and how reliable,
are the one-loop perturbative corrections that we have calculated?
We want to know the answers to these questions for $g^2_{\rm bare}\approx1$,
the value of the lattice coupling constant used in present simulations.
To obtain answers we must choose a value for $\mu a$,
and a scheme for the coupling constant.
We address these two choices in turn.
Throughout this section we refer to the renormalization coefficients
calculated above as ``Z-factors''.

\subsection{Choosing $\mu a$ and $g^2$}

The scalar, tensor and pseudoscalar operators depend upon $\mu a$ because
they have non-vanishing anomalous dimensions.
When these operators appear in the expression for a physical amplitude,
however, they will always be multiplied by a
coefficient function such that the product is independent of $\mu$.
In this sense the choice of $\mu$, and thus $\mu a$, is irrelevant.
In practice, however, the coefficient function is only known to a given order
in perturbation theory, as is the Z-factor, so there is some
dependence on $\mu$, suppressed by powers of $g^2$.
The same applies for the dependence of coefficient functions and
Z-factors on the renormalization scheme.
It is for this reason that we give the results as a function of $\mu a$
and for various schemes.

It is useful to explicitly show how this works. We do so for
the simplest example of an operator which does not mix with others.
A physical amplitude is then given by
\begin{equation}
\ca_{\em phys} \propto C(\mu) \vev{\co_\CONT}_\mu
		= C(\mu) Z(\mu a) \vev{\co_\LATT}_a \ ,
\end{equation}
where the first expectation value indicates a continuum matrix element with
renormalization scale $\mu$,
the second a lattice matrix element with lattice spacing $a$.
The product $C\times Z$ should be independent of $\mu$, and of scheme,
as long as one calculates to all orders in perturbation theory.
We assume that $C(m_H)$ is known, in a particular scheme,
at a heavy scale $m_H$. For example, if the operator is a scalar,
then $C(m_H)$ is the running mass at that scale.
The solution of the renormalization group equation for $C$ is then
(in the notation of Ref. \cite{buras})
\begin{equation}
\label{coeffunction}
C(\mu) = C(m_H)
\left[ g^2(m_H) \over g^2(\mu) \right]^{\gamma^{(0)}/2\beta_0}
\left[ 1 + {g^2(m_H) - g^2(\mu) \over 16 \pi^2}
\big({\gamma^{(1)}\over2\beta_0} - {\gamma^{(0)}\beta_1 \over 2\beta_0^2}\big)
+ O(g^4) \right] \ .
\end{equation}
Here $g(\mu)$ is the running coupling constant in the scheme being
used, while $\beta_n$ and $\gamma^{(n)}$ are the $n$'th order
terms in the beta-function and anomalous dimension, respectively.
If we use another scheme to define the operators, then the form of
the result is the same, but $C(m_H)$ and $\gamma^{(1)}$ would differ.
Given the result in Eq. \ref{coeffunction},
one can show that the Z-factor must take the following form
in order that the product $C \times Z$ be independent of $\mu$:
\begin{eqnarray}
\nonumber
Z(\mu a) &=& 1 + {g^2(\mu)\over 16\pi^2}
\left[ - \gamma^{(0)} \ln(\mu a/\pi) + c \right]
+ \left(g^2(\mu)\over 16\pi^2\right)^2
\left[\half \gamma^{(0)}(\gamma^{(0)}\!-\!2\beta_0) \ln^2(\mu a/\pi)\right. \\
\label{zfactor}
&&\quad \left.\mbox{} -
(\gamma^{(1)}+(\gamma^{(0)}\!-\!2\beta_0)c) \ln(\mu a/\pi)
      + c' \right] + O(g^6) \ .
\end{eqnarray}
This form applies if the same definition is used for the
coupling constant as in Eq. \ref{coeffunction}.
Scheme dependence enters through the constants $c$ and $c'$,
and through $\gamma^{(1)}$.
Our result (Eq. \ref{contlatt}) yields the well known results
$\gamma^{(0)}=-2 C_F d_i$, and also gives the value of $c$
($c=C_F c_{ii}$ for those operators which do not mix).

To obtain the physical amplitude we must
multiply Eqs. \ref{coeffunction} and \ref{zfactor}.
We then see the well known result that the terms multiplying
$g^2(\mu)$ in the second parenthesis in the expression for $C(\mu)$
are of the same order as the $O(g^2)$ corrections to $Z(\mu)$
(``1-loop matching requires 2-loop anomalous dimensions'').
If $g^2(m_H)<< g^2(\mu)$, so that we can ignore terms proportional to
$g^2(m_H)$, then the full first order correction to the leading
logarithms is
\begin{equation}
\label{fullcorrection}
{g^2(\mu)\over 16\pi^2}
\left[ - \gamma^{(0)} \ln(\mu a/\pi) + c +
\gamma^{(0)} \beta_1/2\beta_0^2 - \gamma^{(1)}/2\beta_0 \right] \ .
\end{equation}
This combination must be scheme independent,
as long as we always use the same definition of the coupling constant,
so that the leading logarithmic term in $C(\mu)$ is fixed.
Thus the scheme dependence of $c$,
which enters through the constant $t_S$ as shown in Eq. \ref{mixingcs},
must cancel that in $\gamma^{(1)}/2\beta_0$.\footnote{%
The scheme dependence in $C(m_H)$ will cancel
that of the term proportional to $g^2(m_H)$ in $C(\mu)$.
As an incidental comment we note that our results for $t_S$ for the
vector and axial currents allow us to deduce the two-loop anomalous
dimension for the $V-A$ current in the \HV\ scheme. The result
agrees with that of Ref. \cite{buras}.}
This means that, to evaluate the size of the corrections,
we need to know both $c$ and $\gamma^{(1)}$ in a particular scheme,
in addition to the scheme independent results for
$\beta_0$, $\beta_1$ and $\gamma^{(0)}$.

Unfortunately, while we know $\gamma^{(1)}$ for scalar
and pseudoscalar bilinears
(where the three loop result is given by \cite{kataev}),
and for vector and axial currents
(for which, in \NDR, $\gamma=0$ to all orders),
we do not know the result for the tensor.
For both the scalar and pseudoscalar $\gamma^{(0)}=-6 C_F$, and
$\gamma^{(1)}= -C_F(101-10 N_f/3)$ in the \NDR\ scheme.\footnote{%
We are assuming flavor non-singlet operators,
which differ from singlet operators at the 2-loop level.}
Setting $N_f=0$ (appropriate for the quenched approximation),
and using the quenched values $\beta_0=11$ and $\beta_1=102$,
we find that
$\gamma^{(0)} \beta_1/2\beta_0^2-\gamma^{(1)}/2\beta_0\approx2.8$.
For the tensor the first term alone gives $\approx1.1$.
These numbers are small compared to the values of $c=C_F c_{ii}$
in Table \ref{diagcoefs}, and we choose to ignore them when
discussing the size of the corrections.
This allows us to treat the tensor bilinear,
for which we do not know $\gamma^{(1)}$, in the same way as the others.
We choose to use the values of the $c_{ii}$ in the \DREZ\ scheme,
although the conclusions are essentially unchanged in the other schemes.

As an aside, we note that it follows from the previous discussion that
we can deduce the value of the 2-loop anomalous dimension on the lattice
{}from that in a continuum scheme:
\begin{equation}
\gamma^{(1)}_\LATT = \gamma^{(1)}_\CONT - 2\beta_0 c
+ 2 \beta_0 \gamma^{(0)} \ln(\Lambda_\CONT/\Lambda_\LATT) \ .
\end{equation}
This relationship involves the ratio of $\Lambda$-parameters because the
coupling constant is different in the lattice scheme.
In fact, for the operators under consideration here, we could just
as well dispense with the continuum scheme altogether.
This is not true, however, for the main application of our results.
This is the calculation of Z-factors for the
four-fermion operators which result when the $W$ and $Z$ are
integrated out of the electroweak theory.
Since we cannot put the weak interactions on the lattice,
as the theory is chiral, the coefficients $C(m_H)$ must be computed
in the continuum, and we must match the resulting operators with
those on the lattice.

Now we return to the issue of which value of $\mu a$ to use
when estimating the size of the corrections.
The value should be chosen so that the higher order terms in
$Z$ do not contain large logarithms proportional to $[g^2 \ln(\mu a)]^n$.
To do this we should match the lattice cut-off $\sim \pi/a$ with
the continuum renormalization point $\mu$.
Our choice, as already noted above, is to use $\mu=\pi/a$.
A value of, say, $\mu=1.7\pi/a$ would be equally reasonable,
and would shift the coefficients $c_{ii}$ by $2\ln(1.7) d_i\approx d_i$.
This is smaller than most of the coefficients,
so the ambiguity in $\mu$ is only of minor practical significance.
It is also important to note that the change in $c_{ii}$
depends only on the spin of the operator,
so that a change in $\mu$ does not alter the large difference
between $c_{SS}$ and $c_{SP}$ in column (a) of Table
\ref{diagcoefs}.

We must also choose a definition of the coupling constant
in the expression for $Z$,
and the scale at which to evaluate the coupling, which we call $\mu'$.
Different choices give different higher order constants $c'$.
Lepage and Mackenzie argue that higher order constants are minimized
if one uses a continuum-like scheme such as $\bar{\rm MS}$
\cite{lepagemackenzie}.
They also suggest a prescription for choosing $\mu'$,
which would, in general, give a different result for each operator.
We have chosen to simplify the discussion
by choosing the same scale for all operators,
and taking this to be the same as that in the explicit logarithms:
$\mu'=\mu=\pi/a$.
In fact, this choice has been made already in the expression for the
Z-factor, Eq. \ref{zfactor}.
The result is that, if the bare lattice coupling is $g^2_{\rm bare}\approx1$,
as in present simulations, one should use $g^2(\mu'=\pi/a)\approx1.8$.
Taking this value, the factor by which to multiply the numbers in
Tables \ref{diagcoefs} and \ref{mixingcoefs}
is $1.8\times C_F/(16 \pi^2)\approx 0.015 \approx 1/66$.

\subsection{The size of the corrections}

After this long digression we return to the size of the corrections.
Multiplying the numbers in Tables \ref{diagcoefs} and \ref{mixingcoefs}
by $1/66$,
we see that the corrections for gauge invariant operators (column (a))
vary from about $-\half$ to greater than $\frac23$.
The only exception is the mixing coefficients, which are uniformly small.
The large size and range of the corrections
suggests that there is considerable uncertainty
in the application of perturbation theory to such operators.

A large part of the perturbative corrections, however, comes from
fluctuations in the links: there is a systematic increase in $c_{ii}$
as the number of links increases.
These fluctuations are reduced by using projected links,
as shown in column (b)
by the considerable reduction in the range of corrections.
The Landau gauge bilinears (column (c)) have no links,
and also show a reduced range of corrections,
roughly from $-\half$ to $0$.
This result, together with the fact that these operators are simple to
implement, makes these operators attractive for numerical simulations.

Finally, the smeared bilinears have the smallest range of corrections,
roughly from $-0.2$ to $0$.
There is very little variation between the vector and
tensor operators. In addition, the mixing coefficients are all smaller
than $1\%$. Thus, to a reasonable approximation, these operators are
simply slightly rescaled versions of the corresponding continuum operators.
This makes them good candidates for further numerical study.

\subsection{Summing tadpoles}

Lepage and Mackenzie show that tadpole diagrams are the main source
of the large difference between the bare lattice coupling
and $g^2_{\bar{\rm MS}}$ \cite{lepagemackenzie}. They suggest a mean field
method for removing the dominant effect of tadpole diagrams.
They have applied this to pure gauge theory and to Wilson fermions.
Here we apply their method to staggered fermions, in order to see
how well it reduces the fluctuations in the operators.

The idea is that the trace of gauge links $U_\mu$ fluctuates
around a value $u_0$, rather than 1, because of the tadpole diagrams.
It is the rescaled links $U_\mu/u_0$ which one should expand around unity,
and use in perturbation theory.
In particular, the staggered fermion action
\begin{equation}
S = \sum_n \left( m \bar\chi(n) \chi(n) +
\half \sum_\mu \eta_\mu(n)
\bar\chi(n) [U_\mu(n)\chi(n+\mu)-U_\mu(n-\mu)^{\dag} \chi(n-\mu)] \right)
\ ,
\end{equation}
should be rewritten in terms of $\psi=\sqrt{u_0}\chi$
\begin{eqnarray}
\nonumber
S &=& \sum_n \left( (m/u_0) \bar\psi(n) \psi(n) + \right.\\
&& \left. \half \sum_\mu \eta_\mu(n)
\bar\psi(n) [(U_\mu(n)/u_0)\psi(n+\mu)-
(U_\mu(n-\mu)^{\dag}/u_0) \psi(n-\mu)] \right)
\ .
\end{eqnarray}
The claim is that $\psi$ is better matched to the continuum quark field
than is $\chi$,
and that one should use $\psi$ to construct lattice operators.
Similarly, if such operators contain links, then they should be
the rescaled links $U_\mu/u_0$.
For the bilinears we are studying, the prescription is then to
multiply the naive lattice operator by $u_0^{1-n_U}$,
where $n_U$ is the number of gauge links.
For the gauge invariant operators $n_U=\Delta$,
for the Landau gauge operators $n_U=0$,
while for operators with projected gauge links we assume that
$n_U=1-\delta_{\Delta,0}$.

In addition, we should subtract from the perturbative corrections
$c_{ij}$ the parts coming from the tadpole diagrams.
This is because such diagrams are the lowest order contribution to
$u_0$, and thus have already been taken into account.
The actual value of $u_0$ is to be determined non-perturbatively,
in effect summing up tadpoles (and possibly some other effects) to all
orders. The precise amount to subtract from $c_{ij}$ is weakly
dependent on the definition of $u_0$ used. We opt for the definition
$u_0=\langle\frac13 \Tr(U)_L\rangle$,
the trace of the link in Landau gauge.
In perturbation theory this means that
$u_0=1- C_F (g^2/16\pi^2) 9.17479+ O(g^4)$.
This choice has the advantage that it amounts to dropping from $c_{ii}$
the contribution from all tadpole diagrams (ZT and T) evaluated in Landau
gauge.

The same prescription applied to the gauge action itself
predicts that the correct coupling constant to use in perturbative
formulae is $g^2_{\rm bare} u_0^{-4}$ \cite{lepagemackenzie}.
For $g^2_{\rm bare}=1$, numerical simulations yield
$\langle\frac13 \Tr(U)_L\rangle\approx0.86$.
Taking this as our estimate of $u_0$, we find that the effective
coupling should be $1.8 g^2$, which is the value chosen above.

\begin{table}
\begin{center}
\begin{tabular}{crrrr}
Name &	(a) \quad &(b) \quad &(c) \quad &(d) \quad \\
\hline 
$c_{SS}$  & -22.1825 & -22.1825 & -22.1825 &   0.5074 \\
$c_{SP}$  &  18.7011 &  10.3456 &   9.9523 &  10.5628 \\
$c_{SV}$  &  -0.7743 &  -0.7743 &  -0.7743 &   7.2443 \\
$c_{SA}$  &  14.4948 &  10.3273 &   8.7527 &  10.1605 \\
$c_{ST}$  &   9.1279 &   8.4331 &   6.3620 &   9.3952 \\
$c_{VS}$  &   0.0000 &   0.0000 &   0.0000 &  -1.4538 \\
$c_{VP}$  &   4.1595 &  -0.0080 &  -1.5826 &  -1.9639 \\
$c_{VV0}$ &  -5.6050 &  -5.6050 &  -5.6050 &  -3.4702 \\
$c_{VV2}$ &   0.8659 &   0.1711 &  -1.9000 &  -1.9304 \\
$c_{VA2}$ &   0.8840 &   0.1892 &  -1.8819 &  -2.1511 \\
$c_{VA4}$ &   6.9764 &  -1.3791 &  -1.7724 &  -1.8307 \\
$c_{VT1}$ &  -3.3958 &  -3.3958 &  -3.3958 &  -2.7134 \\
$c_{VT3}$ &   3.8926 &  -0.2749 &  -1.8495 &  -1.8842 \\
$c_{TS}$  &  -0.9987 &  -1.6935 &  -3.7646 &  -5.1562 \\
$c_{TV1}$ &  -2.4711 &  -2.4711 &  -2.4711 &  -4.9461 \\
$c_{TV3}$ &   1.1142 &  -3.0533 &  -4.6279 &  -5.2590 \\
$c_{TT0}$ &   0.5875 &   0.5875 &   0.5875 &  -4.1294 \\
$c_{TT2}$ &  -1.3147 &  -2.0095 &  -4.0806 &  -5.2010 \\
$c_{TT4}$ &   3.7349 &  -4.6207 &  -5.0140 &  -5.2952 \\
\end{tabular}
\end{center}
\caption[taddiagcoefs]{\fc
Results for the diagonal part of renormalization constants at one-loop,
$c_{ii}$, with Landau gauge tadpoles removed, as appropriate in
the mean field prescription of Lepage and Mackenzie.
The notation is as in Table 6. 
}
\label{taddiagcoefs}
\end{table}

To illustrate the effect of this prescription we show
the shifted diagonal corrections in Table \ref{taddiagcoefs}.
Notice that the prescription does not affect the conserved vector
current at all, since $ZT = - T_{\Delta=1}$.
It does affect the Landau gauge version of this current,
shifting its correction to zero.
The result of the shifts is, as expected, most significant for
the gauge invariant operators.
The corrections now range roughly from $-\frac13$ to $\frac13$,
possibly small enough that perturbation theory is reliable.
The range of the corrections is also reduced for the
projected operators, for which the corrections remain
slightly smaller than those for the gauge invariant operators.
The tadpole subtraction simply shifts the corrections for the
Landau gauge operators, with the result that they are
more evenly distributed around zero. The smeared operators
still have the smallest corrections.
Finally, we note that the prescription has no effect
on the off-diagonal corrections,
because tadpoles contribute only to diagonal renormalization.

\section*{ACKNOWLEDGMENTS}

We are grateful to Rajan Gupta and Greg Kilcup for many discussions,
and for participation in the early stages of this work,
and to David Daniel for reading the manuscript.
We thank Narahito Ishizuka, Masanori Okawa, Yoshihisa Shizawa
for comparing results and for comments on the manuscript,
and Roger Horsley for discussions of numerical results.
AP thanks Los Alamos National Laboratory
for hospitality during part of this work.
SS is supported in part by the DOE under contract
DE-AC05-84ER40150 and grant DE-FG09-91ER40614,
and by an Alfred P. Sloan Fellowship.

%

\def\MPA#1#2#3{{Mod. Phys. Lett.} {\bf A#1} (#2) #3}
\def\PRL#1#2#3{{Phys. Rev. Lett.} {\bf #1} (#2) #3 }
\def\PRD#1#2#3{{Phys. Rev.} {\bf D#1} (#2) #3}
\def\PLB#1#2#3{{Phys. Lett.} {\bf #1B} (#2) #3}
\def\NPB#1#2#3{{Nucl. Phys.} {\bf B#1} (#2) #3}
\def\NPBPS#1#2#3{{Nucl. Phys.} {\bf B ({Proc. Suppl.}){#1}} (#2) #3}

\def\etal{{\em et al}}
\def\ringberg#1{Proceedings of the Ringberg Workshop,
	{\sl ``Hadronic Matrix Elements and Weak Decays''},
        Ringberg, Germany, 1988, edited by A. Buras \etal,
	\NPBPS{7A}{1989}{#1}}
\def\capri#1{
  Proceedings of the International Symposium on Lattice Field Theory,
  {\sl ``LATTICE 89''},
  Capri, Italy, 1989, edited by N. Cabibbo \etal,
  \NPBPS{17}{1990}{#1}}
\def\talla#1{
  Proceedings of the International Symposium on Lattice Field Theory,
  {\sl ``LATTICE 90''},
  Tallahassee, Florida, USA, 1990, edited by U. M. Heller \etal,
  \NPBPS{20}{1991}{#1}}
\def\tsukuba#1{
  Proceedings of the International Symposium on Lattice Field Theory,
  {\sl ``LATTICE 91''},
  Tsukuba, Japan, 1991, edited by M. Fukugita \etal,
  \NPBPS{26}{1992}{#1}}
\def\amsterdam{
  Talk at the International Symposium on Lattice Field Theory,
  {\sl ``LATTICE 92''}, Amsterdam, The Netherlands, 1992,
  to be published in Nucl. Phys. {\bf B ({Proc. Suppl.}) } }


\begin{thebibliography}{99}
\bibitem{lusignoli}
	{M.Lusignoli, L. Maiani, G. Martinelli and L. Reina,
	\NPB{369}{1992}{139}}
\bibitem{bkprl}
	{G. Kilcup, S. Sharpe, R. Gupta and A. Patel, \PRL{64}{1990}{25}}
\bibitem{sharpelat91}
	{S. Sharpe,  \tsukuba{197}}
\bibitem{sharatchandra}{H.S. Sharatchandra, H.J. Thun and P. Weisz,
	\NPB{192}{1981}{205}}
\bibitem{goltermansmit}
	{M.F.L. Golterman and J. Smit, \NPB{245}{1984}{61}}
\bibitem{daniel}
	{D. Daniel and S. Sheard, \NPB{302}{1988}{471}}
\bibitem{kluberg}
	{H. Kluberg-Stern, A. Morel, O. Napoly and B. Petersson,
	\NPB{220}{1983}{447}}
\bibitem{horsley}
	{R. Horsley, \amsterdam{}}
\bibitem{tsukuba}
	{N. Ishizuka, \amsterdam{}; \\
	 M. Fukugita, N. Ishizuka, H. Mino, H. Okawa and A. Ukawa,
	 preprint KEK-TH-338}
\bibitem{sheardtad}
	{S. Sheard, Edinburgh preprint 88/431 (January 1988)}
\bibitem{lepagemackenzie}
	{G.P. Lepage and P. Mackenzie, \talla{173};
	 preprint FERMILAB-PUB-19/355-T (9/92); \\
	 G.P. Lepage, \tsukuba{45}}
\bibitem{wius}
	{S. Sharpe, A. Patel, R. Gupta, G. Guralnik and G. Kilcup,
	\NPB{286}{1987}{253}}
\bibitem{hmks}
	{R. Gupta, G. Guralnik, G. Kilcup and S. Sharpe,
	\PRD{43}{1991}{2003}}
\bibitem{sharpelat90}
	{S. Sharpe, \talla{429}}
\bibitem{book}
	{S.~Sharpe, in ``Standard Model, Hadron Phenomenology
	and Weak Decays on the Lattice'',
	Ed. G.~Martinelli, to be published by World Scientific
	[University of Washington preprint, DOE/ER/40614-5 (July 1991)]}
\bibitem{mart4fermion}
	{G. Martinelli, \PLB{141}{1984}{395}}
\bibitem{tsukubapert}
	{N. Ishizuka and Y. Shizawa, in preparation; \\
	 N. Ishizuka in \cite{tsukuba}}
\bibitem{usinprep}
	{A. Patel and S. Sharpe, in preparation}
\bibitem{sheard}
	{S. Sheard, \NPB{314}{1989}{238}}
\bibitem{ssringberg}
	{S. Sharpe, \ringberg{255}}
\bibitem{kieu}
	{D. Daniel and T.D. Kieu, \PLB{175}{1986}{73}}
\bibitem{vandendoel}
	{C. van den Doel and J. Smit, \NPB{228}{1983}{122}}
\bibitem{siegel}
	{W. Siegel, \PLB{84}{1979}{193}}
\bibitem{uclapert}
	{C. Bernard, T. Draper and A. Soni, \PRD{36}{1987}{3224}}
\bibitem{altarelli}
	{G. Altarelli, G. Curci, G. Martinelli and S. Petrarca,
	 \NPB{187}{1981}{461}}
\bibitem{thooft}
	{G. 't Hooft and M. Veltman, \NPB{44}{1972}{189}}
\bibitem{buras}
	{A. Buras and P. Weisz, \NPB{333}{1990}{69}}
\bibitem{verstegen}
	{D. Verstegen, \NPB{249}{1985}{685}}
\bibitem{smitvink}
	{J. Smit and J. Vink, \NPB{298}{1988}{557}}
\bibitem{kataev}
	{S. Gorishny, A. Kataev, S. Larin and L. Surguladze,
	 \PRD{43}{1991}{1633}}
\end{thebibliography}
\end{document}